\documentclass[%
reprint,
showkeys,
superscriptaddress,
nofootinbib,
nobibnotes,
amsmath,amssymb,
prb,
floatfix,
]{revtex4-2}

\usepackage{graphicx}
\usepackage{dcolumn}
\usepackage{bm}
\usepackage{hyperref}
\usepackage[mathlines]{lineno}
\usepackage{xcolor}
\usepackage{orcidlink}

\hypersetup{colorlinks=true,%
            linkbordercolor=teal,
            citecolor=teal,
            urlcolor=teal,
            filecolor=teal,
            linkcolor=teal}

\definecolor{orcidlogocol}{HTML}{A6CE39}

\begin{document}



\title{Moir\'e plane wave expansion model for scanning tunneling microscopy\\simulations of incommensurate two-dimensional materials}%


\author{M. Le Ster \orcidlink{0000-0002-3874-799X}}
\email{maxime.lester@fis.uni.lodz.pl}
\author{P. Dabrowski}
\author{P. Krukowski \orcidlink{0000-0002-1368-6908}}
\author{M. Rogala \orcidlink{0000-0002-7898-5087}}
\author{I. Lutsyk}
\author{W. Ry\'s}
\affiliation{Department of Solid State Physics, Faculty of Physics and Applied Informatics, University of Lodz, Pomorska 149/153, Lodz, 90-236, Poland}

\author{T. M\"arkl}
\author{S. A. Brown}
\affiliation{The MacDiarmid Institute for Advanced Materials and Nanotechnology, School of Physical and Chemical Sciences, Te Kura Mat{\=u}, University of Canterbury, Private Bag 4800, Christchurch 8140, New Zealand}

\author{J. S{\l}awi{\'n}ska \orcidlink{0000-0003-1446-1812}}
\affiliation{Zernike Institute for Advanced Materials, University of Groningen,N, Netherlands}

\author{K. Palot{\'a}s \orcidlink{0000-0002-1914-2901}}
\affiliation{Institute for Solid State Physics and Optics, HUN-REN Wigner Research Center for Physics, H-1525 Budapest, Hungary}

\author{P. J. Kowalczyk \orcidlink{0000-0001-6310-4366}}
\email{pawel.kowalczyk@uni.lodz.pl}
\affiliation{Department of Solid State Physics, Faculty of Physics and Applied Informatics, University of Lodz, Pomorska 149/153, Lodz, 90-236, Poland}

\date{\today}

\begin{abstract}
Incommensurate heterostructures of two-dimensional (2D) materials, despite their attractive electronic behaviour, are challenging to simulate because of the absence of translation symmetry. Experimental investigations of these structures often employ scanning tunneling microscopy (STM), however there is to date no comprehensive theory to simulate an STM image in such systems. In this paper, we present a geometry-based method to generate STM images in incommensurate van der Waals (vdW) heterostructures, which we call the moir\'e plane wave expansion model (MPWEM). We generate the STM images using a weighted sum of three image terms: the non-interacting STM images of (1) the substrate layer, (2) the adsorbate layer, and (3) a semi-empirical Fourier expansion of the moir\'e wavevectors obtained analytically which results from the interaction of (1) and (2). We illustrate and benchmark the model using selected vdW 2D systems composed of structurally and electronically distinct crystals, and show that the method reproduces experimental STM images down to angstrom-scale details. The MPWEM, thanks to its simplicity, can serve as an initial prediction tool prior to more computationally expensive calculations which are often limited by the number of atoms and the requirement of periodic supercells, and thus offers a promising class of computationally-friendly STM simulations, of high relevance in the growing field of twistronics.

\end{abstract}


\maketitle
\newpage

\section{Introduction}

Combining two-dimensional (2D) materials into heterostructures or layering them on the surface of bulk materials is one of the most active fields of condensed-matter physics \cite{Andrei2021}. Electronic coupling of 2D materials often gives rise to moir\'e superlattices because of the alteration of the translational, rotational and mirror symmetries that are present in the individual lattice structures. The impact of the moir\'e superlattices on the electronic, magnetic and optical properties in 2D heterostructures is the subject of an increasing number of studies \cite{Balents2020, Andrei2021, He2021, Mak2022, Jadaun2023, Du2023}, and arguably the archetypal example is that of `magic angle' unconventional superconductivity \cite{Cao2018}, and correlated insulator states \cite{Cao2018b, Su2022} in twisted bilayer graphene. Many other physical properties that have a moir\'e origin have been uncovered, including unconventional ferromagnetism \cite{Sharpe2019} and ferroelectricity \cite{Zheng2020, Kang2023}, moir\'e excitons \cite{Wilson2021}, moir\'e solitons \cite{Kang2021}, topological states \cite{MinPark2021, Liu2021} as well as a wealth of lattice relaxations \cite{Ushida2014, Carr2018, Li2021}. Moir\'e superlattices are also increasingly considered as a tuneable periodic template for atomic \cite{Trishin2021} and cluster adsorption \cite{JimenezSanchez2021}. In general, moir\'e physics can enable access to otherwise elusive quantum states of matter \cite{Kennes2021}, and may provide avenues to experimentally realize Majorana edge modes and topological superconductivity \cite{Kezilebieke2022, Khosravian2024}. Incommensurate systems have long been in the focus of theoretical investigations, with prototypical examples including the Frenkel-Kontorova model \cite{Kontorova1938} (describing the structure of dislocations), the Anderson localization \cite{Anderson1958} (in amorphous structures) and the Aubry-Andr\'e models \cite{Aubry1980} (external periodic potentials incommensurate to atomic lattices) among others \cite{Bak1982}. Moir\'e systems have renewed the significance of incommensurate phenomena such as the Hofstadter fractal energy spectra \cite{Hofstadter1976}, experimentally realized in hBN/graphene \cite{Hunt2013} and bilayer graphene \cite{Lu2021} heterostructures.

Most of the theoretical investigations in moir\'e physics have focused on the origin of flat bands and developing correlation models for systems including ‘magic-angle’ twisted bilayer graphene \cite{Peltonen2018, Haddadi2020, Su2022, Zhou2022}. The field is still in its infancy, and tasks such as predicting an STM image are typically performed using density functional theory (DFT), which can be computationally challenging, in part due to the large supercells that can be required for large moir\'e cells \cite{Carr2020, Pham2022}. The STM calculation method itself may be limiting: the Bardeen's method \cite{Bardeen1961} for example considers all single electron states in the vacuum junction stemming from the sample surface and the STM tip to build the tunneling matrix elements, and is thus extremely time consuming for large systems. Even though efficient DFT-based STM simulation methods have been developed either by neglecting tip effects, such as the Tersoff-Hamann approximation \cite{Tersoff1983, Tersoff1985}, or including the tip electron orbitals and electronic structure, such as the Chen's derivative rules \cite{Chen1990, Mandi2015}, the size of the large (in many cases, still approximate) DFT supercell is difficult to overcome. More importantly, DFT-based STM calculations are clearly unable to treat incommensurate multilayer 2D material systems inherently to the DFT supercell method. Beyond fundamental relevance, the development of a straightforward and computationally undemanding STM simulation method for incommensurate 2D materials is of particular interest in the ever-growing field of vdW heterostructures, where STM is paramount to experimental investigations at the nanoscale. Fast and efficient methods in STM simulation could assist scientists in the identification of the layered 2D crystals and allow for a quick estimate of experimental crystallographic parameters, \emph{e.g.}, the twist angle $\theta$, based on the moir\'e superlattice visible on an atomically resolved image. The simulations could also be performed ahead of time to predict the appearance of the moir\'e superlattice for a given twist angle or as a function of strain; additionally it could assist in the design of heterostructures with the geometry of the superlattice as an input.

In this paper we present a simulation method based on the plane wave description of the non-interacting layers in the heterostructure (i.e., the knowledge of the amplitude and phase shifts of the relevant wavevectors) which we call the moir\'e plane wave expansion model (MPWEM). The simulated STM image is a composite of three weighted terms: (1) the non-interacting substrate STM image, (2) the non-interacting adsorbate STM image, and (3) a Fourier expansion utilizing the moir\'e wavevectors with analytically determined phase shifts, and whose amplitude follow a monotonic dependence on the moir\'e wavevector length. After introducing the MPWEM in detail, we then illustrate its usefulness on a number of vdW heterostructures, showing its potential to generate STM images with fidelity with the observations. Finally, we consider the meaning of its parameters separately, and discuss the limitations of our approach.

\begin{figure*}[t!]
    \centering
    \includegraphics[width=\textwidth]{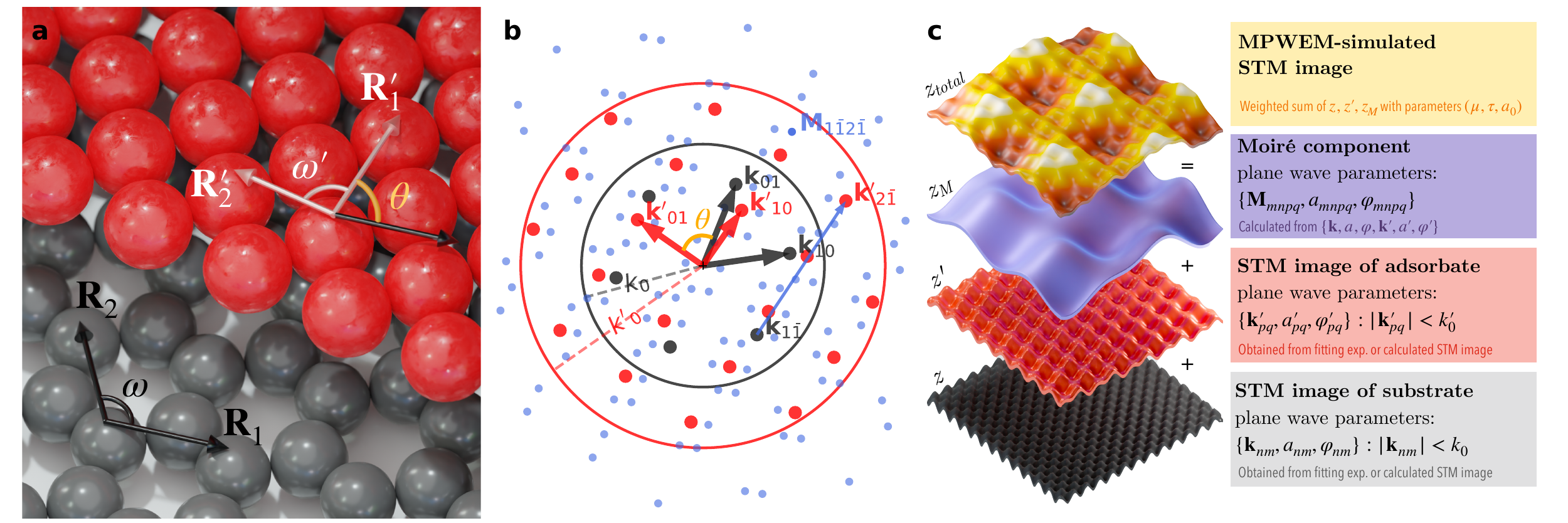}
    \caption{(a) Real space and (b) reciprocal lattices of two arbitrary crystals misaligned by a twist angle $\theta$. The substrate (adsorbate layer) vectors are shown in black (red), the moir\'e wavevectors are shown as blue dots. A moir\'e vector, $\mathbf{M}_{1\bar{1}2\bar{1}}$, is highlighted as an example, equivalent to the blue arrow. The black and red circles indicate the cutoff values $k_0$ and $k'_0$, respectively. (c) Concept of the MPWEM showing the different terms representing the system: the substrate layer ($z$), adsorbate layer ($z'$), moir\'e or interaction term ($z_M$), and total STM image ($z_{total}$). The total STM image is a weighted sum of the other terms.}
    \label{fig:1}
\end{figure*}

\section{Model}

\subsection{Non-interacting substrate and adsorbate}

The MPWEM inputs are the two STM images of the non-interacting system: the substrate and the adsorbate layers, which can either be a 2D material (for both substrate and adsorbate layer) or the surface of a crystalline bulk material or thin film (substrate layer). We define the 2D or surface real space unit cell by two base vectors ($\mathbf{R}_1$ and $\mathbf{R}_2$) spanning an unit cell angle $\omega$ \cite{LeSter2019a, LeSter2019b}, as shown schematically in  Fig~\ref{fig:1}(a). The two lattices are twisted by an angle $\theta$ and may also be translated in plane. Importantly, the two structures are considered rigid, \emph{i.e.}, local lattice distortions and structural defects \cite{Yang2023} are not taken into account. Defect-free STM images of single crystals are periodic and thus we describe the STM images of both non-interacting layers with a Fourier expansion:
\begin{equation}
\label{eq:planewave}
\begin{split}
z(\mathbf{r}) = \sum_{m,n} a_{mn}\cos{(2\pi\mathbf{k}_{mn}\cdot\mathbf{r}+\varphi_{mn})}\\
z'(\mathbf{r}) = \sum_{p,q} a'_{pq}\cos{(2\pi\mathbf{k}'_{pq}\cdot\mathbf{r}+\varphi'_{pq})}
\end{split}
\end{equation}
where $\mathbf{k}_{mn}$ and $\mathbf{k}_{pq}'$ are the reciprocal lattice vectors (note, $\mathbf{k}_{mn}=m\mathbf{k}_{10}+n\mathbf{k}_{01}$) up to their respective cutoff values $k_0$ and $k'_0$ ($|\mathbf{k}_{mn}|<k_0$ and $|\mathbf{k}'_{pq}|<k'_0$), $\varphi$ and $\varphi'$ the phase shifts and $a$, $a'$ the amplitudes ($a, a' \in\mathbb{R}>0$). Increasing $k_0$ and $k'_0$ leads to a larger number of plane waves (growing with the square of the cutoff) and possibly to a more accurate description of the STM image. For convenience, the plane wave amplitudes are normalized, \emph{i.e.},~\mbox{$\sum_i a_i = \sum_j a'_j = 1$} (a scaling prefactor can be used in eq. (\ref{eq:planewave}) to faithfully reproduce the experimental STM image scale). The primitive lattice symmetries are encoded in $\mathbf{k}_{10}$ and $\mathbf{k}_{01}$ ($\mathbf{k}'_{10}$ and $\mathbf{k}'_{01}$) and the detail of the unit cell (related to the electronic density and/or topography) in the amplitude and phase shifts. Because $z$ and $z'$ are known from experiment or simulation, the individual electronic properties of the substrate and the adsorbate layers are contained in the plane wave parameters $\{ (a_i, \varphi_i) \}$ and $\{(a'_j, \varphi'_j) \}$ respectively.

\subsection{Moir\'e contribution}

As shown previously \cite{LeSter2019b}, the moir\'e wavevectors resulting from the coupling of the substrate and adsorbate layers are given by:
\begin{equation}
\label{eq:Mmnpq}
\mathbf{M}_{mnpq} = \mathbf{k}'_{pq} - \mathbf{k}_{mn},
\end{equation}
excluding $(m, n) = (0,0)$ and $(p, q) = (0, 0)$. This results from the convolution theorem in Fourier analysis (see proof of eq. (\ref{eq:Mmnpq}) in Supplementary Information (SI) section \ref{si:convolution}). We also find that the set of $\mathbf{M}_{mnpq}$ within the first Brillouin zone (FBZ) is repeated across other Brillouin zones (see SI section \ref{si:fbz}). An example moir\'e wavevector, $\mathbf{M}_{1\bar{1}2\bar{1}}$ resulting from the hypothetical coupling of $\mathbf{k}_{1\bar{1}}$ and $\mathbf{k}'_{2\bar{1}}$, is highlighted in Fig.~\ref{fig:1}(b) (see also the arrow connecting $\mathbf{k}_{1\bar{1}}$ and $\mathbf{k}'_{2\bar{1}}$ representing the difference vector). The moir\'e or interaction term $z_M$ is also defined as a Fourier expansion as follows:
\begin{equation}
\label{eq:planewave2}
z_M(\mathbf{r}) = \sum_{m, n, p, q} a_{mnpq}\cos{(2\pi\mathbf{M}_{mnpq}\cdot\mathbf{r}+\varphi_{mnpq})},
\end{equation}
where $a_{mnpq}$ and $\varphi_{mnpq}$ are the moir\'e amplitudes and phase shifts. The amplitudes $a_{mnpq}$ are normalized similarly to the non-interacting layer plane wave expansions (again, scaling parameters will control the moir\'e contribution $z_M$ in the final image, which we will introduce in a moment). The fundamental structural difference between the individual non-interacting terms ($z$ and $z'$) and the moir\'e term $z_M$ is that the set of wavevectors $\mathbf{M}_{mnpq}$ does not form a regular lattice in reciprocal space for the general case of incommensurate superposition, analogous to the diffraction patterns of incommensurately modulated bulk crystals \cite{VanSmaalen1995} and quasicrystals \cite{Goldman1993}, and therefore $z_M$ (and the simulated STM image) is quasi-periodic in the general case.

Similarly to the non-interacting layer terms shown in eq. (\ref{eq:planewave}), both amplitudes and phase shifts will play an important role in determining the shape and appearance of the $z_M$ term, and it is therefore crucial to develop a rationale to predict these quantities. The moir\'e phase shifts are obtained by considering two arbitrary plane waves and the resulting phase shifts of the product (see section \ref{si:phaseshifts} in the SI for a more detailed calculation). It follows that:
\begin{equation}
\label{eq:phaseshifts}
\varphi_{mnpq} = \varphi'_{pq} - \varphi_{mn}.
\end{equation}

Despite the Fourier analysis result in section \ref{si:convolution}, which suggests that the moir\'e amplitudes `inherit' the amplitudes from their `parent' amplitudes $a_{mnpq} = a_{mn}a'_{pq}$, this is not what our experimental observations indicate (see SI section \ref{si:amplitudes}). Instead, the moir\'e amplitudes tend to decay as a power law of the norm of their wavevector as follows:
\begin{equation}
 a_{mnpq}=c~|\mathbf{M}_{mnpq}|^{\eta}
 \label{eq:amplitudepowerlaw}
\end{equation}
with $\eta$ the moir\'e damping parameter ($\eta<0$) and $c$ a normalization constant $c\in\mathbb{R}>0$ such that $\sum a_{mnpq} = 1$ and $a_{mnpq}\geq0$. In the investigated systems, $|\eta|\sim 2-3$, as shown in SI section \ref{si:amplitudes} where experimental moir\'e amplitudes are fit to eq. (\ref{eq:amplitudepowerlaw}).

\subsection{Total contributions}

Figure~\ref{fig:1}(c) illustrates the central hypothesis of the MPWEM, that is, the total STM image of the \emph{interacting} system is the weighted sum of three terms: (1) the non-interacting substrate layer $z(\mathbf{r})$, (2) the non-interacting adsorbate layer $z'(\mathbf{r})$, and (3) the moir\'e term (resulting from and representing the interaction) $z_M(\mathbf{r})$. We group both non-interacting terms under the $z_{ni}$ term as follows:
\begin{equation}
\label{eq:z0z1zm}
z_{ni} = (1-\tau)\cdot z + \tau\cdot z'
\end{equation}
where $\tau$ is the opacity parameter ($\tau=1$ means that the signal from the substrate lattice is completely hidden by the adsorbate layer). The non-interacting term combines with the moir\'e term as follows:
\begin{equation}
\label{eq:z0z1zm}
z_{total} = a_0 \times \left[(1-\mu)\cdot z_{ni} + \mu \cdot z_M\right]
\end{equation}
where $a_0$ is a scaling factor (in \AA), $\mu$ a dimensionless moir\'e coupling scalar ($\mu=0$ means that the STM image is dominated by the non-interacting term $z_{ni}$ with a negligible moir\'e interaction term, whereas $\mu=1$ indicates a strong moir\'e coupling such that the non-interacting images are dwarfed by the moir\'e term $z_M$). In other words, eq. (\ref{eq:z0z1zm}) defines the total STM image as a linear combination of two ideal (periodic) STM images and a moir\'e term whose magnitude is tuned by the coupling parameter $\mu$. This is valid under a weakly interacting hypothesis, \emph{i.e.}, where the coupling between substrate and adsorbate layer does not significantly alter the intra-layer crystallography, such that structural and chemical properties (\emph{e.g.} interatomic distances, orbital hybridization) remain preserved upon mild coupling, as expected for vdW heterostructures \cite{Geim2013, Novoselov2016, Wang2021}.

\begin{figure*}[t]
    \centering
	\includegraphics[width=\textwidth]{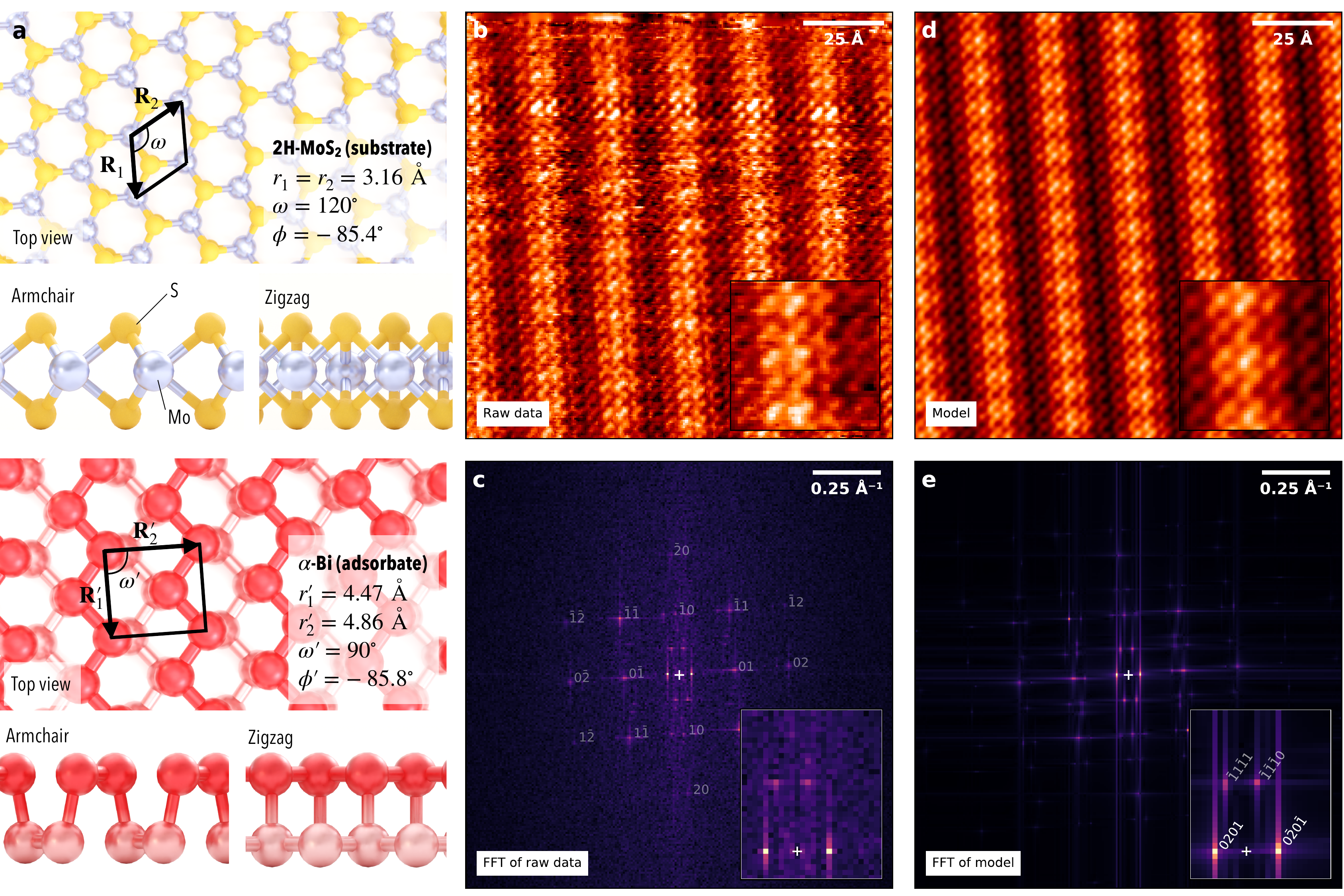}
    \caption{MPWEM on the $\alpha$-Bi/MoS$_2$ system. (a) MoS$_2$ (adsorbate) and $\alpha$-Bi (bottom) crystal structures viewed from top, armchair and zigzag projected views, the lattice parameters and unit cell vectors are indicated. (b) Experimental raw (unfiltered) STM image of $\alpha$-Bi/MoS$_2$ (constant current mode, $V=0.10$~V, $I=10$~pA). (c) Modulus of the FFT of the STM image in (b), the relevant reciprocal lattice points of $\alpha$-Bi are indicated. The other features correspond to the moir\'e wavevectors. (d) Simulated STM image using the MPWEM with the following parameters: $k_0=k'_0=0.48$~\AA$^{-1}$, $\mu=0.64$, $\tau=0.95$ and $\eta=-2.18$. (e) Modulus of the FFT of the simulated STM image in (d). Insets in (b, d) are higher magnifications ($25\times25$~\AA$^2$) at the same $(x,y)$ coordinates. Insets in (c, e) are magnifications near the origin, highlighting the main moir\'e plane waves. Note that the colour scale of FFTs in the entire paper is set to enhance low intensity features.}
    \label{fig:bimos2_main}
\end{figure*}

\section{Results}

We now illustrate the MPWEM using two 2D systems to demonstrate its potential for STM image simulation. We discuss the fidelity of the resulting simulated STM images in each case. We investigate two 2D/surface systems which includes semi-metals (graphene, WTe$_2$), a 2D topological insulator ($\alpha$-Bi), and a semiconductor (MoS$_2$).

\subsection{$\alpha$-Bi on MoS$_2$}
\label{section:bimos2}

\begin{figure*}[t] 
    \centering
	\includegraphics[width=\textwidth]{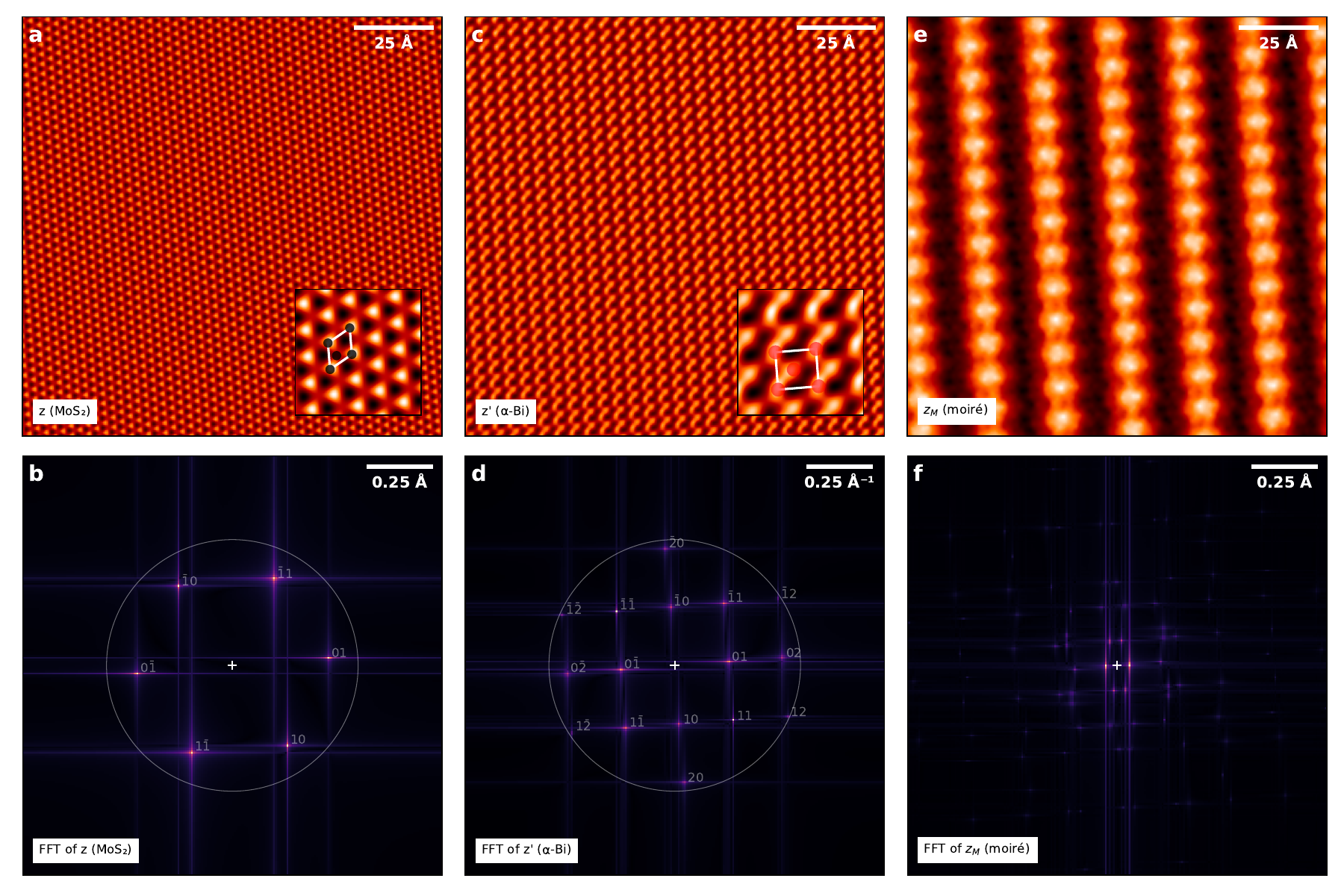}
    \caption{Different terms of the MPWEM for the $\alpha$-Bi/MoS$_2$ system. (a) Substrate layer MoS$_2$'s $z(\mathbf{r})$, rotated with respect to $\mathbf{x}$-axis by $\varphi=-85.4^\circ$. The inset is a higher magnification of $z(\mathbf{r})$, the unit cell is indicated. (b) Modulus of the FFT of $z(\mathbf{r})$, the reciprocal lattice points of MoS$_2$ used in the MPWEM calculation are indicated, as well as the reciprocal cutoff $k_0=0.48$~\AA$^{-1}$ (circle). (c) adsorbate layer $\alpha$-Bi's $z'(\mathbf{r})$, rotated by $\varphi' = -85.8^\circ$ from the $\mathbf{x}$-axis, \emph{i.e.}, $\theta=\varphi'-\varphi = -0.4^\circ$. The inset is a higher magnification of $z'(\mathbf{r})$, with overlaid schematic unit cell. (d) Modulus of the FFT of $z'(\mathbf{r})$ in (c), with reciprocal lattice points $|\mathbf{k'}_{pq}|<k'_0=0.48$~\AA$^{-1}$ indicated (circle has a radius $k'_0$). (e) Moir\'e or interaction term $z_M(\mathbf{r})$ obtained for $\eta=-2.18$, \emph{i.e.}, $a_{mnpq}\propto|\mathbf{M}_{mnpq}|^{-2.18}$. (f) Modulus of the FFT of the moir\'e term in (c).}
    \label{fig:bimos2_detail}
\end{figure*}

\paragraph{{Experimental description.}} The alpha phase of bismuthene ($\alpha$-Bi) is synthesized by physical vapour deposition on the bulk MoS$_2$ substrate, forming ultra thin ($\simeq3$~\AA) islands. As opposed to the wide majority of reported moir\'e structures, this system has the particularity of combining two different Bravais lattices: the transition metal dichalcogenide MoS$_2$ substrate (space group: P6$_3$/mmc) has a surface triangular symmetry and the topological insulator $\alpha$-Bi adsorbate (space group: Pmna \cite{Kowalczyk2020}) has a black phosphorus-like rectangular unit cell (crystal structures and lattice parameters shown in Fig.~\ref{fig:bimos2_main}(a)). The growth details have been published elsewhere \cite{Markl2017}, and the geometry of the moir\'e superlattice in this system has also been separately investigated \cite{LeSter2019b}. The moir\'e superlattice was recently shown \cite{Salehi2023} to modulate the topological edge states of $\alpha$-Bi when deposited on HOPG.

\paragraph{Experimental STM image.} Figure~\ref{fig:bimos2_main}(b) shows an atomically-resolved STM image of $\alpha$-Bi/MoS$_2$. The small-scale protrusions correspond to Bi atoms, and the fringes running nearly vertically across the image correspond to a moir\'e modulation of relatively large amplitude and wavelength. Upon closer look, additional modulations of shorter moir\'e length (and weaker amplitudes) are also resolved. Figure~\ref{fig:bimos2_main}(c) shows the FFT of the STM image shown in (b). The reciprocal lattice of $\alpha$-Bi is well-resolved, as expected from the atomically-resolved STM image (the indices $(p,q)$ are overlaid onto the FFT). In addition the FFT highlights a number of features, in particular in the vicinity of the origin (see inset). These three independent modulations correspond to the moir\'e modulations mentioned above (we discuss their origin later in the text). A series of faint (nonetheless distinct from noise) satellite features further away from the origin are also resolved in proximity of higher order $(p,q)$ reciprocal lattice points.

\paragraph{MPWEM simulation.} We now look at the MPWEM-simulated STM image for this system, shown in Fig.~\ref{fig:bimos2_main}(d) (optimized parameters reported in the caption of Fig.~\ref{fig:bimos2_main}, optimization method is detailed in section SI \ref{si:computationalmethods}). The similarities with the raw data in Fig.~\ref{fig:bimos2_main}(b) are immediately established. The atomic modulations of $\alpha$-Bi is evidently similar because the adsorbate layer term $z'(\mathbf{r})$ corresponding to the $\alpha$-Bi non-interacting term is extracted from an atomically-resolved STM image of $\alpha$-Bi from fitting 2D plane waves (with wavevectors determined by the reciprocal lattice of $\alpha$-Bi, within the reciprocal radius given by $k'_0$; see section \ref{si:computationalmethods} in the SI for method details). The MPWEM successfully predicts the existence of the main moir\'e modulation whose fringes run nearly vertically in the image, as well as the more subtle and more subtle modulations with shorter lengths. The FFT of the MPWEM-simulated STM image in Fig.~\ref{fig:bimos2_main}(e) confirms the agreement (compare with the FFT of experimental STM image in Fig.~\ref{fig:bimos2_main}(c)), where virtually all features in the FFT are reproduced by the MPWEM. The MPWEM-simulated STM image appears to agree extremely well with the experimental data, globally (the presence of correct moir\'e modulations in terms of moir\'e lengths and amplitude), but also on the local scale (the phase shifts $\varphi_M$ are correctly determined). This is highlighted with the higher magnification images shown in insets of Fig.~\ref{fig:bimos2_main}(b, d). The MPWEM is able to capture \AA-scale details, for instance several atomic protrusions appear higher in topography than others in both the experimental and simulated STM images. At first glance, these local higher protrusions in the experimental data may be assigned to local defects (e.g. adsorbed contaminant in the adsorbate layer or structural defect in the substrate layer), yet such local distortions in the topography can be fully understood from a moir\'e origin. Note, that the higher magnification insets are cropped from the images at the exact same $(x,y)$ locations, further confirming that the MPWEM parameters are correctly determined.

\paragraph{MPWEM terms.} We now describe the three terms of the MPWEM simulation in more detail. Figure~\ref{fig:bimos2_detail}(a) shows the MoS$_2$ substrate layer used in the simulation above. As shown in the FFT in Fig.~\ref{fig:bimos2_detail}(b), $z(\mathbf{r})$ consists of three independent plane waves only $|\mathbf{k}_{mn}|<k_0=0.48$~\AA$^{-1}$ (the reciprocal cutoff is represented with the circle of radius $k_0$). In the absence of high quality STM images of MoS$_2$, the data shown here is generated using a simple procedure: each of the sublattice (Mo and S) is associated with a triangular lattice consisting of a sum of three biased 2D cosines raised to an exponent (to obtain more or less narrow effective localized radii), weighted, and then added together. The artificial STM image is then fit with 2D cosines with wavevectors up to $|\mathbf{k}|<k_0$. Section \ref{si:computationalmethods} explains this procedure in more detail. Here, for the generation of $z(\mathbf{r})$ we use literature values of lattice parameters ($r_1 = r_2 = 3.15$~\AA, $\omega=120^\circ$, with sublattice sites at $(0,~0)$ and $(2/3,~1/3)$, the latter corresponding to the Mo atoms located below the top S layer) and a Mo/S weight ratio of $0.5$. The inset of Fig.~\ref{fig:bimos2_detail}(a) shows the real space image in higher resolution, which looks indeed very similar to high quality, atomically-resolved pristine STM images of MoS$_2$ obtained at $V=0.1$~V \cite{Addou2015}. The real space $z'(\mathbf{r})$ of the $\alpha$-Bi adsorbate layer, shown in Fig.~\ref{fig:bimos2_detail}(c) is obtained differently, \emph{i.e.}, by fitting 2D plane waves (of wavevectors $|\mathbf{k}'_{pq}|<k'_0=0.48$~\AA$^{-1}$) onto an experimental STM image of $\alpha$-Bi. The choice of this cutoff allows to capture reciprocal lattice points up to $\mathbf{k}'_{12}$, as shown in the FFT in Fig.~\ref{fig:bimos2_detail}(d) (more details on the choice of the reciprocal cutoff are discussed in section \ref{si:recicutoff}). The reason behind these two separate fitting steps is to acquire plane wave parameters for each considered reciprocal lattice point, \emph{i.e.}, the wavevector $\mathbf{k}$ (2D coordinates, $k_x$ and $k_y$), the phase shift $\varphi$ and the real amplitude $a$, required to generate the set of $\mathbf{M}_{mnpq}$. The plane wave fitting method acting on STM images is described in more detail in SI section \ref{si:computationalmethods}. Now that both the substrate and adsorbate layers have a set of plane waves that fully describe their low frequency (below the reciprocal cutoffs) structures, we can perform the calculation of the moir\'e or interaction term $z_M(\mathbf{r})$ using eqs. (\ref{eq:Mmnpq}, \ref{eq:planewave2}, \ref{eq:phaseshifts} and \ref{eq:amplitudepowerlaw}). The result is shown in Fig.~\ref{fig:bimos2_detail}(e), as expected the main moir\'e modulations (with the largest moir\'e lengths) correspond to $\mathbf{M}_{0201}$, $\mathbf{M}_{\bar{1}1\bar{1}1}$ and $\mathbf{M}_{\bar{1}\bar{1}\bar{1}0}$ located in reciprocal space in the vicinity of the origin (see also insets in Fig.~\ref{fig:bimos2_main}(b,d)). Interestingly, due to the absence of commensurability in this system (no reciprocal lattice points of substrate and adsorbate layers share the same reciprocal coordinates, $\mathbf{k'}_{pq}\neq\mathbf{k}_{mn}$, even beyond $k_0$), the resulting moir\'e term $z_M$ is not strictly periodic (see how local maxima are not identical across the $z_M$ image). Additional discussion on the nature of the $\mathbf{M}_{mnpq}$ set with respect to commensurability can be found in SI section \ref{si:commensurability}. The total MPWEM-simulated STM image, previously shown in Fig.~\ref{fig:bimos2_main}(c) is simply a weighted sum of the three terms in Fig.~\ref{fig:bimos2_detail}(a, c, e), as given by eq. (\ref{eq:z0z1zm}). The MPWEM parameters in the summation are, for the dimensionless moir\'e coupling parameter, $\mu=0.64$ (\emph{i.e.}, 64\% of the amplitudes correspond to the $z_M$ term, $z$ and $z'$ share the remaining 36\%), the adsorbate layer opacity parameter $\tau=0.95$ (the non-interacting terms $z$ and $z'$ are 95\% dominated by the adsorbate layer, \emph{i.e.}, only 5\% of the individual lattices corresponds to MoS$_2$), and scaled by $a_0 = 0.42$~\AA. These numbers are obtained using linear regression (see more details in section \ref{si:computationalmethods}).


\subsection{Graphene on WTe$_2$}
\label{section:gwte2}

\begin{figure*}[t] 
    \centering
	\includegraphics[width=\textwidth]{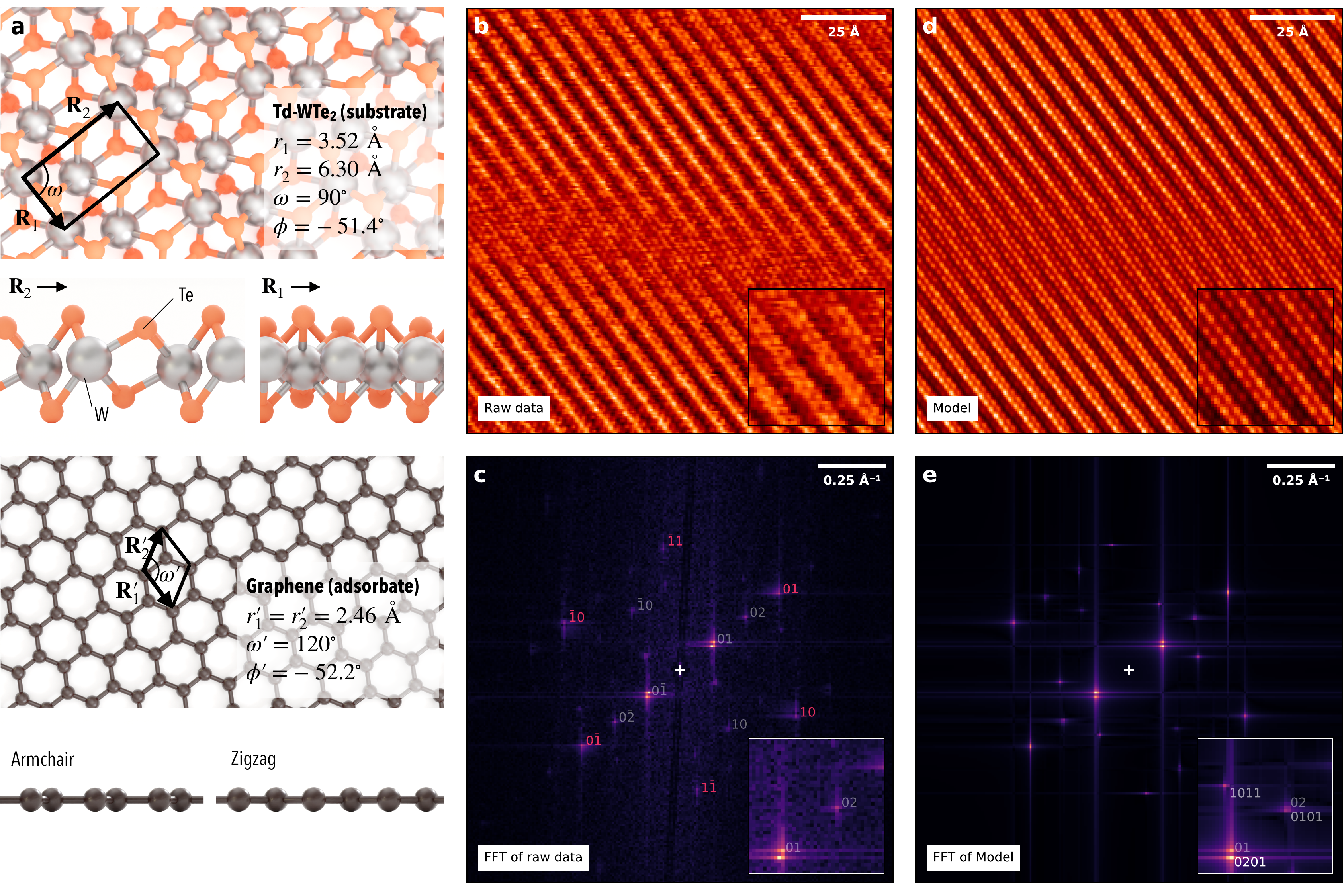}
    \caption{MPWEM on the graphene/WTe$_2$ system. (a) WTe$_2$ (adsorbate) and graphene (bottom) crystal structures viewed from top, armchair and zigzag projected views, the lattice parameters and unit cell vectors are indicated. (b) Experimental raw (unfiltered) STM image of graphene/WTe$_2$ (constant current mode, $V= 0.001$~V, $I=1.5$~nA). (c) Modulus of the FFT of the STM image in (b), the relevant reciprocal lattice points of WTe$_2$ (grey) and graphene (red) are indicated. The other features are identified as moir\'e wavevectors. (d) Simulated STM image using the MPWEM with the following parameters: $k_0=0.32$~\AA$^{-1}$, $k'_0=0.48$~\AA$^{-1}$, $\mu=0.49$, $\tau=0.60$~, and $\eta=-3.16$. (e) Modulus of the FFT of the simulated STM image in (d). Insets in (b, d) are higher magnifications ($25\times25$~\AA$^{2}$) at the same ($x,y$) coordinates. Insets in (c, e) are magnifications near the origin highlighting the main moir\'e modulation.}
    \label{fig:gwte2_main}
\end{figure*}

\begin{figure*}[t] 
    \centering
	\includegraphics[width=\textwidth]{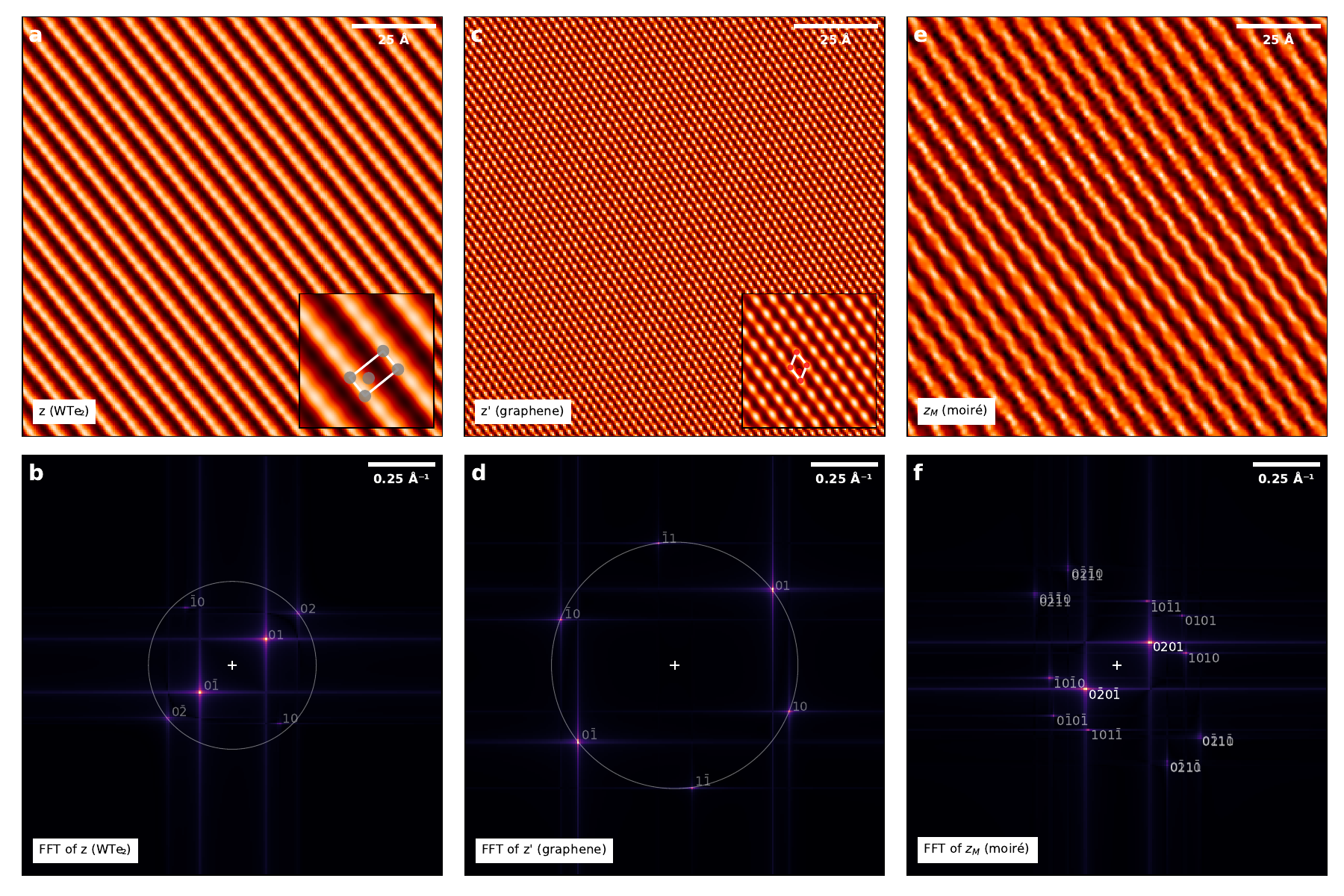}
    \caption{Different terms of the MPWEM for the graphene/WTe$_2$ system. (a) Substrate layer WTe$_2$'s $z(\mathbf{r})$ rotated with respect to the $\mathbf{x}$-axis by $\varphi=-51.4^\circ$. (b) Modulus of the FFT of $z(\mathbf{r})$. (c) adsorbate layer graphene's $z'(\mathbf{r})$, rotated by $\varphi=-52.2^\circ$ from the $\mathbf{x}$-axis, \emph{i.e.}, $\theta=\varphi'-\varphi=-0.8^\circ$. The insets in (a, c) are higher magnifications with the unit cell indicated. (d) Modulus of the FFT of $z'(\mathbf{r})$ in (c), with reciprocal lattice points $|\mathbf{k}'_{pq}|<k'_0$. The circles in (b, d) indicate the reciprocal cutoff values for each layer, $k_0=0.32$~\AA$^{-1}$ and $k'_0=0.48$~\AA$^{-1}$. (e) Moir\'e or interaction term $z_M(\mathbf{r})$ obtained for $\eta=-3.16$, \emph{i.e.}, $a_M\propto|\mathbf{M}|^{-3.16}$. (f) Modulus of the FFT of the moir\'e term in (c) with $(m,n,p,q)$ indices indicated.}
    \label{fig:gwte2_details}
\end{figure*}

\paragraph{Experimental description and STM image.} We now consider the case of monolayer graphene on bulk WTe$_2$ surface in the distorted Td phase \cite{Lee2015}. This time, graphene was transferred onto the substrate using a wet transfer method. Similar to $\alpha$-Bi/MoS$_2$, the two crystal structures possess different symmetries (see Fig.~\ref{fig:gwte2_main}(a) along with the lattice parameters). Figure~\ref{fig:gwte2_main}(b) shows a typical atomically-resolved STM image. Both graphene (hexagonal symmetry, space group P6/mmm) and WTe$_2$ (orthorhombic, space group: Pmn2$_1$) are atomically resolved. The latter is particularly visible as lines propagating from the top left to the bottom right of the screen. The moir\'e structure is different from the case of $\alpha$-Bi/MoS$_2$; instead of an additional modulation on top of graphene, the moir\'e appears to locally damp the modulation caused by the WTe$_2$ atomic rows, in a $\sim20$~\AA-wide region running nearly horizontal across the image. The FFT shown in Fig.~\ref{fig:gwte2_main}(c) confirms these statements: the reciprocal lattices of both graphene and WTe$_2$ are visible, and the additional modulations are attributed to result form the moir\'e interaction. A modulation of particular relevance here, is the FFT feature resolved in the very close vicinity of the reciprocal lattice point $\mathbf{k}_{01}$ (see inset). The presence of two very similar wavevectors in the same image is the cause of the local periodic damping of the WTe$_2$ atomic rows in real space.

\paragraph{MPWEM simulation.} We now investigate this interaction with the MPWEM. The simulated STM image is shown in Fig.~\ref{fig:gwte2_main}(d). Like in $\alpha$-Bi/MoS$_2$ above, the MPWEM allows to generate an STM image with excellent agreement with the observations. The area of local damping of the WTe$_2$ atomic rows is also well reproduced in the MPWEM, in terms of shape, size and effect. The higher magnifications in the insets (extracted from the same coordinates) also indicate a very good \AA-scale agreement, visually confirming the accuracy of the MPWEM parameters. In particular, the graphene atomic sites located on top of the WTe$_2$ modulations (the latter forming nearly continuous stripes), are very well reproduced in the MPWEM. The FFT of the simulated STM image, in Fig.~\ref{fig:gwte2_main}(e), also confirms that most moir\'e waves present in the experimental data are reproduced by the model (insets confirms this further, in particular $\mathbf{M}_{0201}$ is very close to $\mathbf{k'}_{01}$, as discussed previously).

\paragraph{MPWEM terms.} Figure~\ref{fig:gwte2_details} shows the different terms of the MPWEM involved in the making of the simulated STM image in Fig.~\ref{fig:gwte2_main}(c) (optimized parameters in the caption of Fig.~\ref{fig:gwte2_main}). The substrate layer $z(\mathbf{r})$ in Fig.~\ref{fig:gwte2_details}(a) was obtained from extracting the relevant plane wave parameters from the STM image via fitting, and is typical of WTe$_2$ STM images \cite{Tang2017, Peng2017}. The inset shows the unit cell of WTe$_2$, overlaid on top of a magnified region of $z(\mathbf{r})$. The reciprocal cutoff for this layer is $k_0=0.32$~\AA$^{-1}$, set so as to include reciprocal lattice points (wavevectors) up to $\mathbf{k}_{02}$. Note that the experimental STM image also resolves WTe$_2$ in Fig.~\ref{fig:gwte2_main}(a) up to this reciprocal lattice point (beyond $|\mathbf{k}|>k_0$, no feature can be related to the WTe$_2$ lattice). Figure~\ref{fig:gwte2_details}(b) shows the graphene $z'(\mathbf{r})$, also extracted from the STM image, with $|\mathbf{k}'_{pq}|<k'_0 = 0.48$~\AA$^{-1}$ (\emph{i.e.}, only first order reciprocal lattice points). Finally, the moir\'e term $z_M$ resulting from the interaction between $z$ and $z'$ is shown in Fig.~\ref{fig:gwte2_details}(e), calculated with $\eta = -3.16$. The interaction term has a very stripe-like nature, with fringes running nearly (but not exactly) parallel to the WTe$_2$ main atomic rows. These fringes correspond to the $\mathbf{M}_{0201}$ wavevector. Weaker modulations, further away from the origin in reciprocal space, also contribute to the total moir\'e term such as $\mathbf{M}_{1010}$, located in very close vicinity to $\mathbf{k}'_{02}$ (see inset in Fig.~\ref{fig:gwte2_main}(d)). The three MPWEM terms, blended together with the different parameters $\mu=0.49$ and $\tau = 0.60$, lead to the simulated total STM image previously shown in Fig.~\ref{fig:gwte2_main}(c).

\section{Discussion}

\subsection{General applicability of the MPWEM}

The MPWEM proves to be an efficient framework to explain virtually every moir\'e features present in the vdW heterostructures investigated in this paper. The agreement and level of detail that the MPWEM-simulated real space STM images yield is a sufficient proof that the model successfully includes the effect of the moir\'e interaction in atomic-resolution topography STM imaging, and therefore that the interlayer coupling leading to the formation of moir\'e superlattices has an interference origin. The superposition hypothesis (\emph{i.e.}, that the STM image is a sum of two independent terms and one interaction term: the non-interacting (1) substrate and (2) adsorbate images, to which is added (3) the moir\'e term containing the effect of the interaction) is therefore valid. This successfully accounts for the moir\'e contribution in a variety of 2D heterostructures. The specific electronic properties of the layers that couple do not seem to correlate with the MPWEM applicability, since the MPWEM is satisfactory, whether the layer is a semi-metal (graphene), topological semi-metal (WTe$_2$), topological insulator ($\alpha$-Bi) or semiconductor (MoS$_2$). The vdW assembly method (by physical vapour deposition or layer transfer) and the lattice symmetries do not seem to play a role in the validity of the \mbox{MPWEM}. The validity of the model in the presence of stronger interactions between the substrate and the adsorbate layer (\emph{e.g.}, graphene on metallic surfaces \cite{JimenezSanchez2021}) remains to be investigated. We believe that in this case, hybridization may become significant and that the rigid lattice hypothesis may not hold (in this case, the individual STM image components depend on the coupling causing lattice reconstruction). It is worth emphasizing that here, lattice reconstructions were not accounted for, however the agreement shows that the hypothetical atomic displacement which may have occurred in reality was sufficiently weak to be ignored. In the case of low-angle twisted homo- \cite{Carr2019, Haddadi2020, Xie2023, Nakatsuji2023} or heterostructures \cite{Geng2023} (not investigated in this report), accounting for relaxation may be required. Even if a stronger interaction between the substrate and adsorbate layer exists, the separate STM images serving as inputs for MPWEM could be calculated based on the relaxed atomic heterostructure with DFT or tight binding methods.

\begin{table*}[t]
\centering
\begin{tabular}{c|cccccc}

 System & ~~~~$k_0$ (\AA$^{-1}$)~~~~ & ~~~~$k_0'$ (\AA$^{-1}$)~~~~ & ~~~~$\mu$~~~~ & $\tau$ & $\eta$ & $a_0$ (\AA) \\
 \hline\hline
$\alpha$-Bi/MoS$_2$~~~ &~~~~0.48~~~~& ~~~~0.48~~~~& ~~~~0.64~~~~  & ~~~~0.95~~~~ & ~~~~-2.18~~~~ &~~~~0.42~~~~\\
Graphene/WTe$_2$~~~ &~~~~0.32~~~~&~~~~0.48~~~~&~~~~0.49~~~~ & ~~~~0.60~~~~ & ~~~~-3.16~~~~ & ~~~~0.65~~~~\\
\hline

\end{tabular}
\caption{Optimized MPWEM parameters for the different systems.}
\label{table1}
\end{table*}

\subsection{Computation time and DFT}

We stress that the use of a limited number of plane waves for each non-interacting layers (3 independent plane waves for graphene, MoS$_2$ and WTe$_2$; 8 for $\alpha$-Bi), makes the calculation time of the interaction $z_M$ term negligible ($N\times M$ plane waves in the interaction terms, with $N$ and $M$ the number of independent plane waves describing the substrate and adsorbate layer). The fitting procedure, extracting the plane wave parameters (whether from an experimental or simulated STM image) is of the order of seconds, up to several minutes for a large ($>15$) number of independent plane waves. The fitting of moir\'e amplitudes in an experimental image (in the case where the optimal MPWEM parameters $\eta$, $\tau$ and $\mu$ are to be obtained by fitting) may also take several minutes. These characteristic computation times are many orders of magnitude lower than for the standard of STM simulation based on DFT. The determination of the electronic structure by DFT may take up to several weeks of computation power, using a very large unit cell approximating the incommensurate superposition, and STM simulations might take considerable further computational time depending on the level of employed electron tunneling theory \cite{Bardeen1961, Tersoff1983, Tersoff1985, Chen1990, Brandbyge2002, Palotas2014, Mandi2015}. It is worth insisting that our method does not attempt to `replace' DFT as a whole, but simply to serve as a rapid STM simulation method for incommensurate (or commensurate with a very large supercell) vdW system. For a complete physical insight, DFT and other computational methods are paramount.

\subsection{Incommensurability}

The plane wave based method allows to circumvent periodic requirements (as typical in DFT-based STM simulation) and can thus deal with incommensurate systems; the systems investigated in the paper are not specifically designed at commensurate twist angles. It is worth noting that due to the computational nature of the model (through the use of floating point numbers), all the systems investigated here are, strictly speaking, commensurate; however one would require arbitrarily high indices $(n,m)$ and $(p,q)$ in order to satisfy the commensurability condition in reciprocal space, $\mathbf{k}_{mn}=\mathbf{k'}_{pq}$. We explore the consequence of commensurability with respect to the set of $\mathbf{M}_{mnpq}$ in SI section \ref{si:commensurability}.

\subsection{Interpretation}

We treat the $z(\mathbf{r})$, $z'(\mathbf{r})$ and $z_M(\mathbf{r})$ terms as contributions toward a total topographic map, $z_{total}(\mathbf{r})$, which undeniably possesses a topographic quality in accordance to its agreement with experimental STM topography data ($z$ as current map in electric current units in constant height mode, or $z$ as apparent height map in constant current mode of STM). However, in our model the STM imaging mode is not important, and the summing terms remain \emph{dimensionless} quantities expressed in a Fourier expansion sum (of which the amplitudes are normalized), subsequently scaled to topographic scales. The topographic quantity representing the substrate layer is $a_0(1-\mu)(1-\tau)\cdot z$, the adsorbate layer $a_0(1-\mu)\tau\cdot z'$ and the moir\'e term $a_0\mu\cdot z_M$. However, the atomically-resolved image of a surface is not an intrinsic property of the material; instead it depends on a combination of intrinsic and instrumental parameters; \emph{i.e.}, the electronic states, or in a compact representation the local density of states distribution $\rho(\mathbf{r}, E)$, temperature $T$, as well as on a number of STM parameters, mainly, the bias voltage $V$, the tip-sample distance $d$ (which depends on both $V$ and the set-point current $I$ or on altogether specific imaging modes) and finally on the tip shape, symmetry and electronic states with the specific orbital characters at the tip apex which contribute to quantum tunneling. For this reason the MPWEM can not be as complete as a DFT-based STM simulation which can include advanced tunneling theories \cite{Bardeen1961, Tersoff1983, Tersoff1985, Chen1990, Brandbyge2002, Palotas2014, Mandi2015} and can investigate the role of each of these parameters independently. In the MPWEM we can make the assumption that all these different intrinsic and extrinsic parameters are contained in the individual amplitude and phase shifts, as well as in the MPWEM parameters $\mu, \tau$ and $a_0$. Also, $z$ and $z'$ should be obtained at the same bias voltage (shifted to account for charge transfer) in order to generate a valid $z_M$ term, since the electronic densities of states are more likely to couple when at identical energies due to electronic hybridization. The impact of intrinsic parameters (\emph{e.g.} the twist angle $\theta$) and extrinsic parameters on the MPWEM parameters will be explored in a follow-up article.

\subsection{MPWEM parameters}

We now discuss the meaning of the MPWEM parameters; their optimized values in the different systems investigated in this paper are gathered in Table~\ref{table1}.

\paragraph{Reciprocal cutoff.} The role of the cutoff parameters $k_0$ and $k'_0$ is to exclude reciprocal lattice points beyond a certain distance from the reciprocal space origin. Without $k_0$ and $k_0'$, the set of $\mathbf{M}_{mnpq}$ would be infinite (see SI section \ref{si:commensurability}). In the cases considered, the reciprocal cutoffs were set so as to only include the features resolved in the FFT of experimental STM images; beyond which the features are indistinguishable from noise). We investigate the impact of the reciprocal cutoff for the case of $\alpha$-Bi/MoS$_2$ in SI section \ref{si:recicutoff}, in particular the consequences of the under- and overestimation. 

\paragraph{Moir\'e coupling.} The moir\'e coupling parameter $\mu$ is a direct handle on the weight of the interaction in the total STM image. In the investigated cases, the moir\'e coupling parameters are in the vicinity of $\mu\simeq0.5$, indicating that the interaction term and the individual layers contribute equally to the overall modulations in the total image. This quantity is almost certainly expected to vary with both intrinsic and extrinsic parameters. For instance, it is known that the interlayer coupling in twisted bilayer graphene strongly depends on the twist angle \cite{Carr2019}, the moir\'e superlattice is strongly resolved for low twist angles ($\theta\simeq1^\circ$), whereas the layers are decoupled for large twist angles ($\theta > 10^\circ$), corresponding to large and small moir\'e coupling parameters, respectively. Additionally, moir\'e effects occur mostly at energies close to the Fermi level, in other words the moir\'e modulations are visible at low bias voltages (our experimental STM images in this paper were obtained for small biases).

\paragraph{Adsorbate layer opacity.} The adsorbate layer opacity parameter $\tau$ directly controls the visibility of the substrate layer in the total STM image with respect to the adsorbate layer. In this paper, $\tau$ is $0.6$ to nearly 1.0, indicating that most (or nearly all) the individual lattices resolved have an adsorbate layer origin. It is not a surprising result, since STM is a surface sensitive technique where the tunneling current has an exponential dependence on the tip-sample distance $I\sim e^{-\kappa z}$, with $\kappa$ a constant ($\kappa\simeq1$~\AA$^{-1}$). In a simple model where the tunneling current only depends on the tip-sample distance, contributions from the adsorbate and substrate layers are $I'=e^{-\kappa z_0}$ and $I=e^{-\kappa (z_0+h+d_{vdw})}$ respectively, with $z_0$ is the tip-to-top-layer distance, $h$ the adsorbate layer thickness, and $d_{vdw}$ the vdW gap. It follows that $\tau$ can be seen as the ratio of quantum tunneling contribution from the adsorbate layer over the sum of both adsorbate and substrate currents, giving:
\begin{equation}
\tau  = \frac{1}{1+e^{-\kappa (h+d_{vwh})}},
\label{eq:tau}
\end{equation}
indicating that $0.5\leq\tau\leq1.0$. In the $\alpha$-Bi/MoS$_2$ case, the MoS$_2$ lattice is very weakly resolved ($\tau=0.95$) compared to WTe$_2$ in the graphene/WTe$_2$ case ($\tau=0.60$). This is consistent with this tunneling model, where the presence of a larger $\alpha$-Bi thickness, (compared to a single layer of carbon atoms in the second case) reduces the contribution of the substrate layer. Using eq. (\ref{eq:tau}), we obtain (for $\kappa=1$), an additional tunneling thickness of $\Delta z \simeq2.5$~\AA~for the $\alpha$-Bi/MoS$_2$ case compared to graphene/WTe$_2$, which is a relatively good estimate of the additional tunneling distance in this case ($\alpha$-Bi is $\simeq 3$~\AA-thick). The value of $\tau$ is expected to depend on the interplay of the set-point current and bias voltage, since these two determine the tip-sample distance.

\paragraph{Moir\'e damping.} The moir\'e damping parameter $\eta$ acts as a spatial frequency cutoff on the set of moir\'e plane waves. The data suggests that the moir\'e amplitudes seem to depend only on the magnitude of the moir\'e wavevector $\mathbf{M}$, and not on the amplitudes of the parent reciprocal lattice plane waves. The overall damping as a function of the moir\'e vector is expected as the coupling between two electronic plane waves across the vdW junction is likely to have a resonant behaviour leading to larger amplitude of modulation when the detuning $\delta \mathbf{k}$ is small. In a topographic picture, the local quasiperiodic stress caused by the moir\'e superlattice is also more likely to induce an effective topographic corrugation for large moir\'e lengths considering the elastic energy required for the vertical displacement. Lastly, the sharp features in the surface (associated with large $|\mathbf{k}|$) are also damped by a non-ideal blunt tip, due to convolution of the broad apex with the sample features.

\subsection{Future applications and development}

An extended MPWEM model can be developed for three-layer systems (and beyond), with added reciprocal cutoff $k_0^{(i)}$ (for the $i^{th}$ layer), additional layer opacity parameters $\tau^{(i)}$ and moir\'e coupling $\mu^{(i)}$. Moir\'e terms $z_M$ may be considered only for nearest layers or second-nearest, which impacts the number of interaction terms in the system. Second order `moir\'e of moir\'e' terms (where $\mathbf{M}^{(i)}$ and $\mathbf{M}^{(j)}$ interfere together) could also be introduced \cite{Wong2015, Wang2019, Zhu2020, Mao2023}.

Having demonstrated the success of the model to generate the interacting STM image from non-interacting ones, the next natural step in its development is to incorporate energy-dependent information. This will be obtained through assembling scanning tunneling spectroscopy (STS, obtained from experimental STS images or calculated LDOS maps) of the non-interacting layers and test whether a good approximation of the interacting system can be obtained using the MPWEM scheme. This will allow to obtain insight into the role of the bias voltage and set-points towards the total STM image in vdW systems, and to assess energy-dependent MPWEM parameters.

Finally, because of the ability of the MPWEM to encapsulate the moir\'e coupling in STM images, we believe that the MPWEM parameters can serve as a basis to describe moir\'e interactions in any vdW system, provided that the rigid lattice approximation holds (typically valid in large twist angle systems or in low-symmetry van der Waals systems as studied in this paper). Utilizing the MPWEM in a large data approach, with machine learning models \cite{Liu2022} driven by error minimization is still an hypothetical application, but may provide an avenue for future moir\'e-based STM predictive simulation, such that the MPWEM parameters can be correlated with other properties of the 2D system.


\section{Conclusion}

In conclusion, we introduce the MPWEM, a plane wave-based model that successfully captures moir\'e features in incommensurate STM images. Using as an input the two non-interacting images (more essentially, the plane-wave parameters that describe them) and a small number of semi-empirical parameters (describing the intensity of coupling, the adsorbate layer opacity with respect to tunneling and the moir\'e wavevector damping), the total generated STM image possesses high-fidelity with the experimentally obtained STM images of the interacting system. The simplicity and efficiency of calculation of STM images according to the MPWEM makes it a choice model for future simulations of incommensurate 2D van der Waals systems, which were previously not strictly possible based on first-principles calculations. The MPWEM also shows promise for the classification of STM images, boiling down interlayer coupling physics into a handful of parameters.

\section{Methods}

\paragraph{Sample preparation.} The Bi/MoS$_2$ sample was obtained by molecular beam epitaxy in UHV. The Bi solid source (purity $5N$) was evaporated using a standard Knudsen cell, while maintaining the MoS$_2$ substrate at room temperature (target coverage $\sim~1$~nm). The MoS$_2$ substrate was freshly cleaved and degassed for at least an hour at around 150$^\circ$C. The graphene monolayer was obtained by CVD on Ge(001)/Si(100), and later mechanically wet-transferred using a spin-coated PMMA (4\% PMMA solution in anisole) transfer layer. The WTe$_2$ substrate (HQ Graphene) covered with graphene monolayer was then transported to a glovebox, followed by heating and cleaning with acetone.

\paragraph{STM.} The Bi/MoS$_2$ sample was imaged by STM (Omicron VT-AFM) directly after growth in UHV without exposing the sample to ambient air. The graphene/WTe$_2$ sample was transferred directly after preparation from the glovebox into the STM chamber via a portable vacuum suitcase (base pressure $\sim 10^{-8}$ mbar). STM acquisition was performed at room temperature, using manually cut Pt/Ir tips. Raw STM images were drift-corrected after acquisition using our recent FFT-based method \cite{LeSter2024} using MoS$_2$ and graphene's lattice as reference.

\paragraph{Simulations.} The simulated STM images were generated using a python script and a homemade python library (example file accessible at {\url{https://github.com/maximelester/mpwem}}), relying on plane wave parameters. Fitting data to 2D plane waves is obtained with the {\fontfamily{lmtt}\selectfont scipy.optimize.curve\_fit} method and display routines using the {\fontfamily{lmtt}\selectfont matplotlib} library.


\section{Statements}

\paragraph{Data and code availability.} An example python script ($\alpha$-Bi/MoS$_2$ from section \ref{section:bimos2}) with detailed and commented steps for STM data generation is available on GitHub at {\url{https://github.com/maximelester/mpwem}}. The data and code can also be available upon reasonable request to the corresponding authors.

\paragraph{Author Contributions.} M.L.S. conceived and designed the project, led the collaboration, carried out the development of the model, performed the calculations, analysis and programming, and wrote the manuscript. M.L.S., I.L., W.R., T.M and S.A.B. obtained the experimental STM images. P.D., P.K., M.R., J.S., K.P. and P.J.K. assisted M.L.S. with the development of the model, concept, and writing of the manuscript.

\paragraph{Competing Interests.} The authors declare no competing financial interests.

\paragraph{Acknowledgments.} This work was supported by the National Science Center, Poland (M.L.S.: 2022/47/D/ST3/03216; P.J.K., I.L., W.R: 2018/31/B/ST3/02450; P.D.: 2018/30/E/ST5/00667) and IDUB 6/JRR/2021 (M.L.S). J.S. acknowledges the Rosalind Franklin Fellowship from the University of Groningen. K.P. acknowledges the National Research Development and Innovation Office of Hungary (NKFIH, Grant No. FK124100), the J{\'a}nos Bolyai Research Grant of the Hungarian Academy of Sciences (Grant No. BO/292/21/11) and the New National Excellence Program of the Ministry for Culture and Innovation from NKFIH Fund (Grant No. {\'U}NKP-23-5-BME-12). S.A.B. acknowledges the Marsden Fund and the MacDiarmid Institute.

\bibliography{moire}


\clearpage

\newpage
\setcounter{section}{0}
\setcounter{figure}{0}
\setcounter{table}{0}
\renewcommand\thesection{S\arabic{section}}
\renewcommand\thefigure{S\arabic{figure}}
\renewcommand\thetable{S\arabic{table}}

{\centering \large \textbf{Supplementary Information}}
\vspace{2mm}
\hrule
\vspace{2mm}
{\centering \large Moir\'e plane wave expansion model for scanning tunneling microscopy simulations of incommensurate two-dimensional materials}
\vspace{4mm}
\hrule
\vspace{2mm}
{\centering M. Le Ster \emph{et al.}, 2024}


\section{Convolution}
\label{si:convolution}

Let $z$ and $z'$ be the substrate and the adsorbate layer's real space plane wave superposition. The interaction term originates from the product overlap of both layer's charge densities, expressed as Fourier expansion terms. We are then interested in the Fourier transform of $z\times z'$. The convolution theorem in Fourier theory states
\begin{equation}
    \mathcal{F}\{z \times z' \} = \mathcal{F}\{z\} \circledast \mathcal{F}\{z'\}
\end{equation}
where $\circledast$ is the convolution operator, $\mathcal{F}\{z\}$ and $\mathcal{F}\{z'\}$ are the Fourier transforms of $z$ and $z'$, respectively; they are functions of reciprocal space coordinates $\mathbf{k}$. We can then write the convolution as follows:
\begin{equation}
\label{eq-convolution}
    \mathcal{F}\{z \times z'\}(\mathbf{k}) = \int_{-\infty}^{\infty} \mathcal{F}\{z\}(\mathbf{q}-\mathbf{k}) \times \mathcal{F}\{z'\}(\mathbf{q})~d\mathbf{q}.
\end{equation}

The Fourier transform of $z$ and $z'$ are of the form:
\begin{equation}
\label{fakir}
\left\{
    \begin{array}{ll}
    \mathcal{F}\{z\}(\mathbf{k}) = \sum_m \sum_n \alpha_{mn}~ \delta(\mathbf{k}-m\mathbf{k}_{10}-n\mathbf{k}_{01}) \\
    \\
   \mathcal{F}\{z'\}(\mathbf{k}) = \sum_p \sum_q \alpha'_{pq}~ \delta(\mathbf{k}-p\mathbf{k}'_{10}-q\mathbf{k}'_{01})
    \end{array}\right.
\end{equation}
where $\delta(\mathbf{k})$ is the Dirac delta function and $a_{mn}$ complex amplitudes which tend to 0 as $|m|,|n|\rightarrow\infty$. $\mathcal{F}\{z\}(\mathbf{k})=0$ almost everywhere, except when $\mathbf{k}$ is the $(m,n)$ point of the reciprocal lattice in which case $\mathcal{F}\{z\}(\mathbf{k})=\alpha_{mn}$. In other words, the eq. (\ref{fakir}) can be seen as a 2D Dirac comb where impulses are located at the reciprocal lattice points and of complex amplitudes $\alpha_{mn}=a_{mn}e^{i\varphi_{mn}}$ and $\alpha'_{pq}=a'_{pq}e^{i\varphi'_{pq}}$. Inserting (\ref{fakir}) into (\ref{eq-convolution}) gives:
\begin{widetext}
\begin{equation}
\label{big-eq}
    \mathcal{F}\{z\times z'\}(\mathbf{k}) = \int_{-\infty}^{\infty} \left(\sum_m \sum_n \alpha_{mn}~ \delta(\mathbf{q}-\mathbf{k}-m\mathbf{k}_{10}-n\mathbf{k}_{01}) \right)
    \times
    \left(\sum_p \sum_q \alpha'_{pq}~ \delta(\mathbf{q}-p\mathbf{k}'_{10}-q\mathbf{k}'_{01}) \right) d\mathbf{q}
\end{equation}
\end{widetext}
%
%
for which the product inside the integral is zero unless both Dirac delta functions are equal to $1$, \emph{i.e.}, $\mathbf{q}$ satisfies:
\begin{equation}
\label{eq-q}
\left\{
    \begin{array}{ll}
    \mathbf{q}=\mathbf{k} + m\mathbf{k}_{10} + n\mathbf{k}_{01} \\
    \\
    \mathbf{q}=p\mathbf{k}'_{10} + q\mathbf{k}'_{01}
    \end{array}\right.
\end{equation}

The condition in eq. (\ref{eq-q}) is satisfied only if:\begin{equation}
    \mathbf{k} = p\mathbf{k}'_{10} + q\mathbf{k}'_{01} - (m\mathbf{k}_{10} + n\mathbf{k}_{01}).
\end{equation}

We define $\mathbf{M}$ as the difference vector:
\begin{equation}
\label{mnpq_def}
    \mathbf{M}_{mnpq} \triangleq \mathbf{k}'_{pq}-\mathbf{k}_{mn}.
\end{equation}

Finally we rewrite eq. (\ref{big-eq}) as:
\begin{equation}
    \mathcal{F}\{z \times z'\}(\mathbf{k}) = \sum_m \sum_n \sum_p \sum_q \alpha_{mn}~\alpha'_{pq}~\delta(\mathbf{k}-\mathbf{M}_{mnpq})
\end{equation}
which is equivalent to a 2D Dirac comb-like aperiodic lattice determined by $\mathbf{M}_{mnpq}$, with complex amplitudes $\alpha_{M}=\alpha_{mn}\alpha'_{pq}$ (\emph{i.e.} a real amplitude $a_{M}=a_{mn}a'_{pq}$).


\section{Brillouin Zones}
\label{si:fbz}

\begin{figure}[t]
  \centering
  \includegraphics[width=0.666\columnwidth]{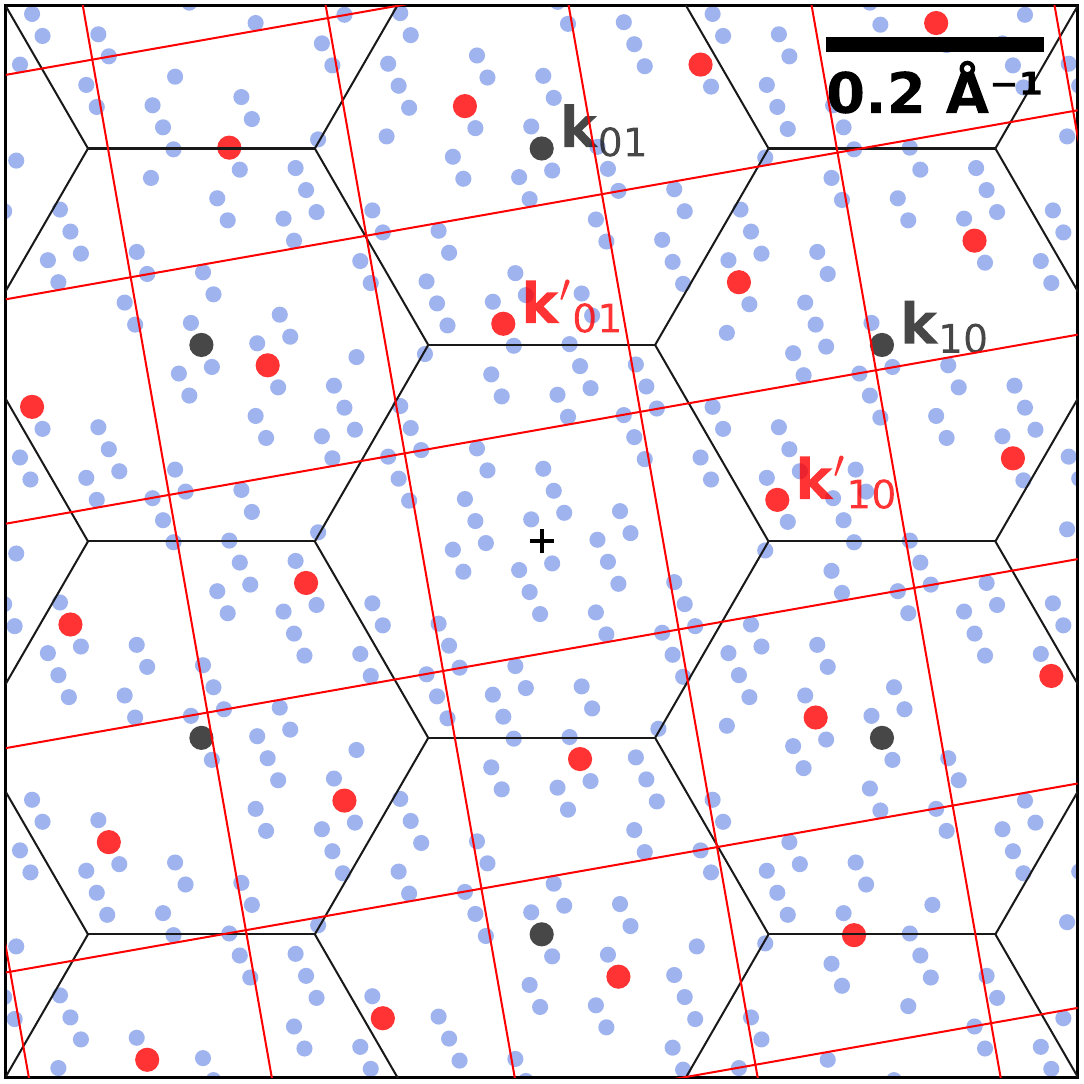}
    \caption{Moir\'e wavevectors and Brillouin zones. Reciprocal lattices of the $\alpha$-Bi/MoS$_2$ system for $\theta=10^\circ$ and $k_0=k_0'=1.0$~\AA$^{-1}$. MoS$_2$ and $\alpha$-Bi reciprocal lattice points and Brillouin zones (centred at the reciprocal lattice points) are indicated in black and red, respectively. The resulting moir\'e wavevector $\mathbf{M}_{mnpq}$ coordinates are shown in light blue dots. The set of $\mathbf{M}_{mnpq}$ is repeated across all Brillouin zones (whether $\alpha$-Bi or MoS$_2$), albeit decreasing in density with increasing $\mathbf{k}$.}
    \label{fig:s3}
\end{figure}

We now investigate the set of moir\'e wavevectors $\mathbf{M}_{mnpq}$ with respect to the Brillouin zones (separated by a reciprocal lattice vector). Let us assume an arbitrary $\mathbf{M}_{mnpq}$ with fixed $m, n, p, q$ indices. We look for alternative indices $m', n', p', q'$ such that:
\begin{equation}
\label{eq:fbz}
\mathbf{M}_{m'n'p'q'} = \mathbf{M}_{mnpq} + \mathbf{k}_{uv}
\end{equation}
where $\mathbf{k}_{uv}$ is the $(u, v)$ reciprocal lattice point of the substrate layer ($u, v\in\mathbb{Z}$). Developing eq. ($\ref{eq:fbz}$), we obtain:
\begin{equation}
\mathbf{M}_{m'n'p'q'} = p\mathbf{k}'_{10}+q\mathbf{k}'_{01} - (m-u)\mathbf{k}_{10} - (n-v)\mathbf{k}_{01}
\end{equation}
hence $(m',n',p',q') = (m-u, n-v, p, q)$. The same demonstration can be made for a translation by an adsorbate layer reciprocal lattice vector $\mathbf{k}'_{uv}$, in which case the indices are $(m',n',p',q')=(m,n,p-u, q-v)$, satisfying the moir\'e wavevector definition in eq. (\ref{mnpq_def}). Figure \ref{fig:s3} shows the reciprocal lattices of an arbitrary system made of $\alpha$-Bi/MoS$_2$, twisted by $\theta=10^\circ$, with reciprocal cutoff $k_0=k'_0=1.0$~\AA$^{-1}$. The nature of the 2D system is not crucial here; we use this as an illustration due to its low symmetry (incommensurate rectangular on hexagonal lattice). The Brillouin zones of MoS$_2$ and $\alpha$-Bi (as well as the reciprocal lattice) are indicated in black and red, and the set of $\mathbf{M}_{mnpq}$ are shown in light blue. As demonstrated rigorously in this section, the set of $\mathbf{M}_{mnpq}$ is repeated across all the Brillouin zones, except decreasing in number per Brillouin zone as $\mathbf{k}$ increases, the first Brillouin zone being the most populated. The decrease of $\mathbf{M}_{mnpq}$ density at larger $\mathbf{k}$ is a consequence of the finite reciprocal cutoff.


\section{Moir\'e phase shifts}
\label{si:phaseshifts}

Let $z$ and $z'$ be arbitrary plane waves of the form $z(\mathbf{r}) = a \cos(2\pi (\mathbf{k}\cdot\mathbf{r})+\varphi)$. We investigate the product $z \times z'$ as follows:
\begin{equation} \label{eq:phase}
\begin{split}
z\times z'  = aa'&\cos(2\pi(\mathbf{k}\cdot\mathbf{r})+\varphi)\cdot\cos(2\pi(\mathbf{k}'\cdot\mathbf{r})+\varphi') \\
  = \frac{aa'}{2} [ &\cos(2\pi\{\mathbf{k'-k}\}\cdot\mathbf{r}+\{\varphi'-\varphi\}) \\
  + &\cos(2\pi\{\mathbf{k'+k}\}\cdot\mathbf{r}+\{\varphi'+\varphi\}) ]
\end{split}
\end{equation}
where the first term in eq. (\ref{eq:phase}) corresponds to our definition of a moir\'e wavevector ($\mathbf{M}=\mathbf{k}'-\mathbf{k}$ in eq. (\ref{mnpq_def})), indicating that $\varphi_M = \varphi'-\varphi$. The second term in eq. (\ref{eq:phase}) is also in fact taken into account in our model, where $\mathbf{k}$ corresponds to $-\mathbf{k}$ (opposite or \emph{twin} reciprocal lattice point with opposite phase).


\section{Moir\'e amplitudes}
\label{si:amplitudes}

\begin{figure*}[t]
  \centering
  \includegraphics[width=0.666\textwidth]{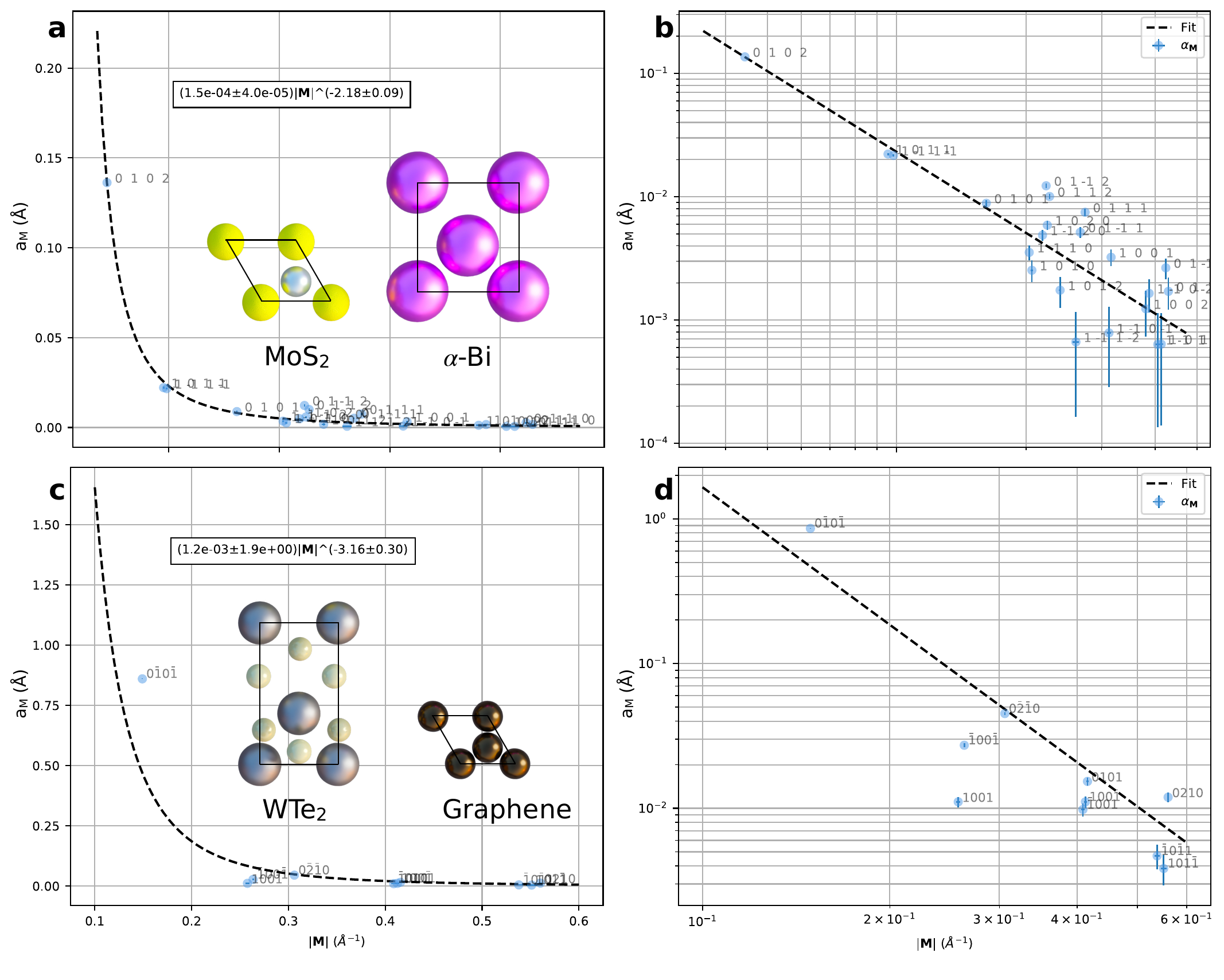}
    \caption{Moir\'e amplitudes $\tilde{a}_{mnpq}$ (not normalized) obtained from fitting plane waves on the STM data. (a-b) $\alpha$-Bi/MoS$_2$ (see section \ref{section:bimos2}). (c-d) Graphene/WTe$_2$ (see section \ref{section:gwte2}). Dashed lines indicate the line of best fit using a power law model. Surface unit cells of the different layers are shown in (a, c). Plots in (a, c) are linear plots and (b, d) are log-log plots.}
    \label{fig:s1}
\end{figure*}

In this section we consider the experimental moir\'e amplitudes in real STM images. We first focus on the $\alpha$-Bi/MoS$_2$ moir\'e amplitudes. Figures \ref{fig:s1}(a) and \ref{fig:s1}(b) show the moir\'e amplitude $a_M$ as a function of $|\mathbf{M}_{mnpq}|$ extracted by fitting plane waves onto the data of the form $a_M \cos(2\pi\mathbf{M}\cdot\mathbf{r} + \varphi_M)$ using a simple least-square algorithm, {\fontfamily{lmtt}\selectfont scipy.optimize.curve\_fit()}. The resulting $a_M = f(|\mathbf{M}|)$ are then fitted with a function of the form $a_M = c \times |\mathbf{M}|^{\eta}$ ($c\in\mathbb{R}$) using the same fitting engine (black line). As shown in the figure, the moir\'e amplitudes indeed follow a power law of exponent $\eta=-2.18$. 

Figure \ref{fig:s1}(b) shows the same method applied to the graphene/WTe$_2$ STM image previously studied in section \ref{section:gwte2}. Generally, the trend is comparable; overall the moir\'e amplitudes decrease with the norm of the wavevector and agree with a power law trend. This time, the power law exponent $\eta$ is slightly larger (in absolute value) with $\eta=-3.16$, indicative of a larger damping of moir\'e amplitudes at larger $|\mathbf{M}|$.

We can not however rule out the potential for other amplitude models, such as an exponential decay \mbox{$a_M \propto e^{\eta}|\mathbf{M}|$} or gaussian centred at origin \mbox{$a_M \propto e^{-{|\mathbf{M}|^{2}}/(2\eta^{2})}$}. Further analyses of STM images of 2D/2D or 2D/bulk coupled systems with clearly resolved moir\'e superlattices, obtained at different twist angles, may provide a deeper insight into the origin of the moir\'e amplitudes.

\section{Choice of reciprocal cutoff}
\label{si:recicutoff}

\begin{figure*}[t]
  \centering
  \includegraphics[width=0.75\textwidth]{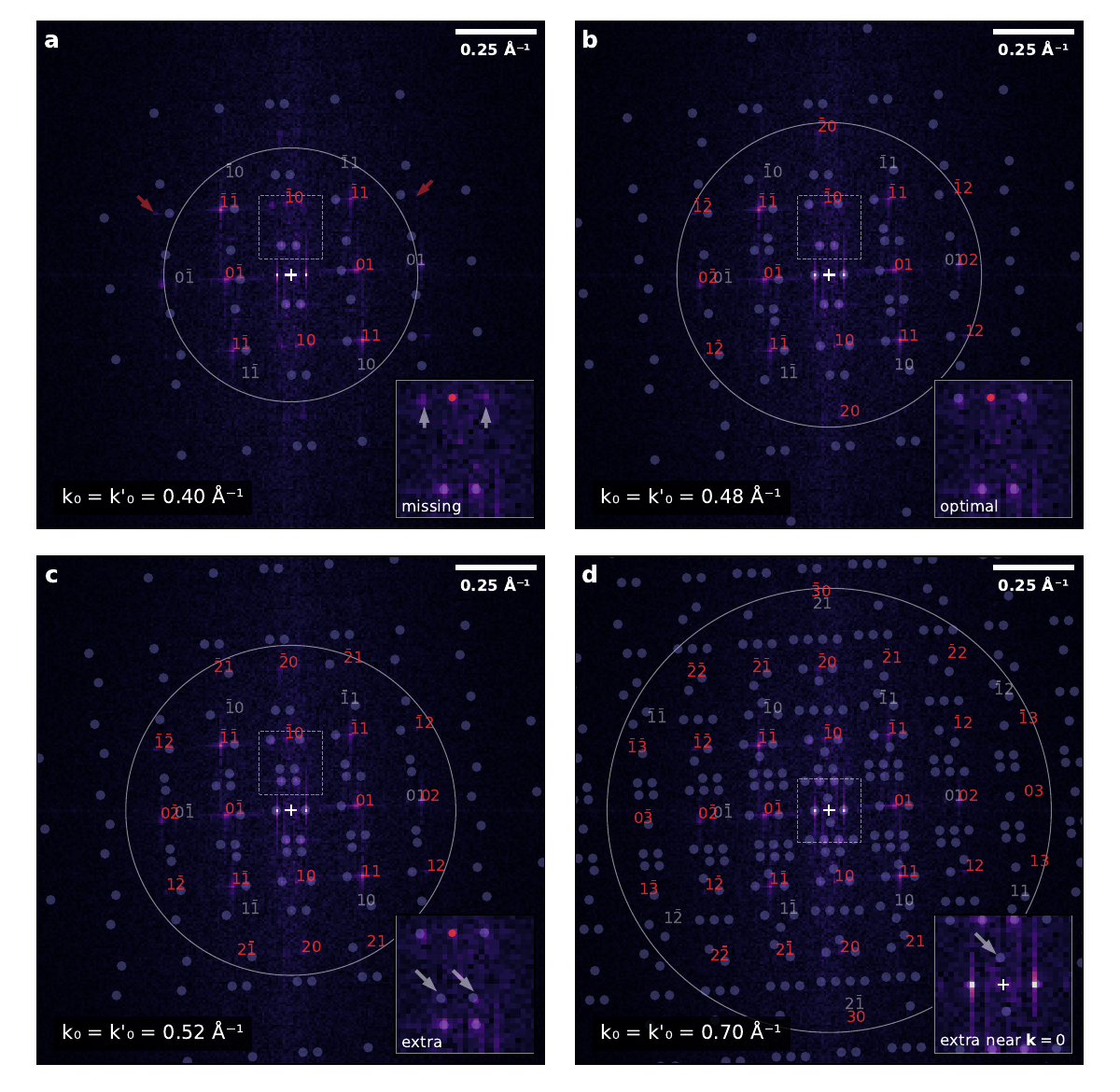}
    \caption{Reciprocal cutoff and MPWEM fidelity. Modulus of the FFT of the STM image shown in section \ref{section:bimos2} in the main text, onto which the predicted moir\'e wavevectors $\mathbf{M}_{mnpq}$ coordinates are overlaid (semi-transparent blue), for (a) $k_0=k_0'=0.40$~\AA$^{-1}$, (b) $k_0=k_0' = 0.48$~\AA$^{-1}$, (c) $k_0=k_0' = 0.52$~\AA$^{-1}$, and (d) $k_0=k_0' = 0.70$~\AA$^{-1}$. The red indices refer to $\mathbf{k'}_{pq}$ (\emph{i.e.}, $\alpha$-Bi) and the grey indices refer to $\mathbf{k}_{mn}$ (MoS$_2$). To avoid clutter, only the location of the predicted $\mathbf{M}_{mnpq}$ is shown as dots, indices are only discussed in the text when relevant. Insets in (a-d) are magnified portions of the FFT delimited by the white dashed squares. Circles centred on origin correspond to the reciprocal cutoffs. The reciprocal cut off is (a) underestimated (several visible moir\'e features are not predicted by the MPWEM, see arrow in inset of (a)), (b) optimal (most moir\'e features are predicted), (c) overestimated (extra $\mathbf{M}_{mnpq}$ predicted by the MPWEM that are not visible in the data), and (d) largely overestimated (new $\mathbf{M}_{mnpq}$ predicted very close to the origin, not visible in the data).}
    \label{fig:si_k0}
\end{figure*}

We now discuss the role of the reciprocal cutoff $k_0$ in terms of agreement with the STM data. Figure~\ref{fig:si_k0}(a) shows the FFT of the STM image previously discussed in section~\ref{section:bimos2}, \emph{i.e.}, $\alpha$-Bi/MoS$_2$. The semi-transparent blue dots are the predicted moir\'e wavevectors obtained with a smaller cutoff,  $k_0=k_0'=0.40$~\AA$^{-1}$, \emph{i.e.} only reciprocal lattice points $\mathbf{k}_{nm}$ and $\mathbf{k'}_{pq}$ contained within the circle of radius $k_0=0.45$~\AA$^{-1}$ will be considered in the MPWEM calculation, slightly lower than the reciprocal cutoff used in the main text ($k_0=k_0'=0.48$~\AA$^{-1}$). First, we consider the case where the reciprocal cutoff is underestimated, shown in Fig.~\ref{fig:si_k0}(a). The MPWEM prediction of the moir\'e wavevectors (24 moir\'e wavevectors) agrees very well with the features present in the FFT of the experimental STM image, however the two satellite features near $\mathbf{k'}_{\bar{1}0}$ (and, by definition, near $\mathbf{k}_{10}$) which are clearly resolved in the FFT (see arrows in the inset in Fig.~\ref{fig:si_k0}(a)), do not have a MPWEM-predicted moir\'e wavevector counterpart. These missing features near $\mathbf{k}_{\bar{1}0}$ ($\mathbf{k}_{10}$) are in fact $\mathbf{M}_{0\bar{1}\bar{1}\bar{2}}$ and $\mathbf{M}_{01\bar{1}2}$ ($\mathbf{M}_{0112}$ and $\mathbf{M}_{0\bar{1}1\bar{2}}$). Their absence among the set of predicted $\mathbf{M}_{mnpq}$ is due to the reciprocal cutoff being too small, excluding $\alpha$-Bi's $\mathbf{k'}_{\bar{1}{2}}$ and $\mathbf{k'}_{\bar{1}2}$ from the MPWEM calculation (see red arrows, pointing at the location of these two reciprocal lattice points sitting slightly outside the reciprocal cutoff circle).

Figure~\ref{fig:si_k0}(b) shows the MPWEM prediction with $k_0=k'_0=0.48$~\AA$^{-1}$, as in the main text. The two moir\'e features which could not be predicted with a lower reciprocal cutoff now have a $\mathbf{M}_{mnpq}$ counterpart (see inset). Increasing slightly the cutoff allowed to include the $\mathbf{k'}_{12}$, $\mathbf{k'}_{\bar{1}2}$, $\mathbf{k'}_{\bar{1}\bar{2}}$ and $\mathbf{k'}_{1\bar{2}}$ reciprocal lattice points of $\alpha$-Bi (indices shown in red in the FFT at their respective location) (total number of $\mathbf{M}_{mnpq}$ is 48).

We now consider the situation where the reciprocal cutoff is slightly overestimated. Figure~\ref{fig:si_k0}(c) shows the same MPWEM for a slighlty larger cutoff: $k_0=k'_0 = 0.52$~\AA$^{-1}$ (60 $\mathbf{M}_{mnpq}$ wavevectors). The reciprocal lattices $\mathbf{k'}_{\bar{2}1}$ and $\mathbf{k'}_{\bar{2}\bar{1}}$ (and the opposite reciprocal lattice points) are now included, which in turn leads to the prediction of additional $\mathbf{M}_{mnpq}$; crucially, in regions where there are no moir\'e features in the data (see arrows in the inset, pointing to $\mathbf{M}_{\bar{1}0\bar{2}\bar{1}}$ and $\mathbf{M}_{\bar{1}1\bar{2}1}$).

\begin{figure*}[t]
  \centering
  \includegraphics[width=0.666\textwidth]{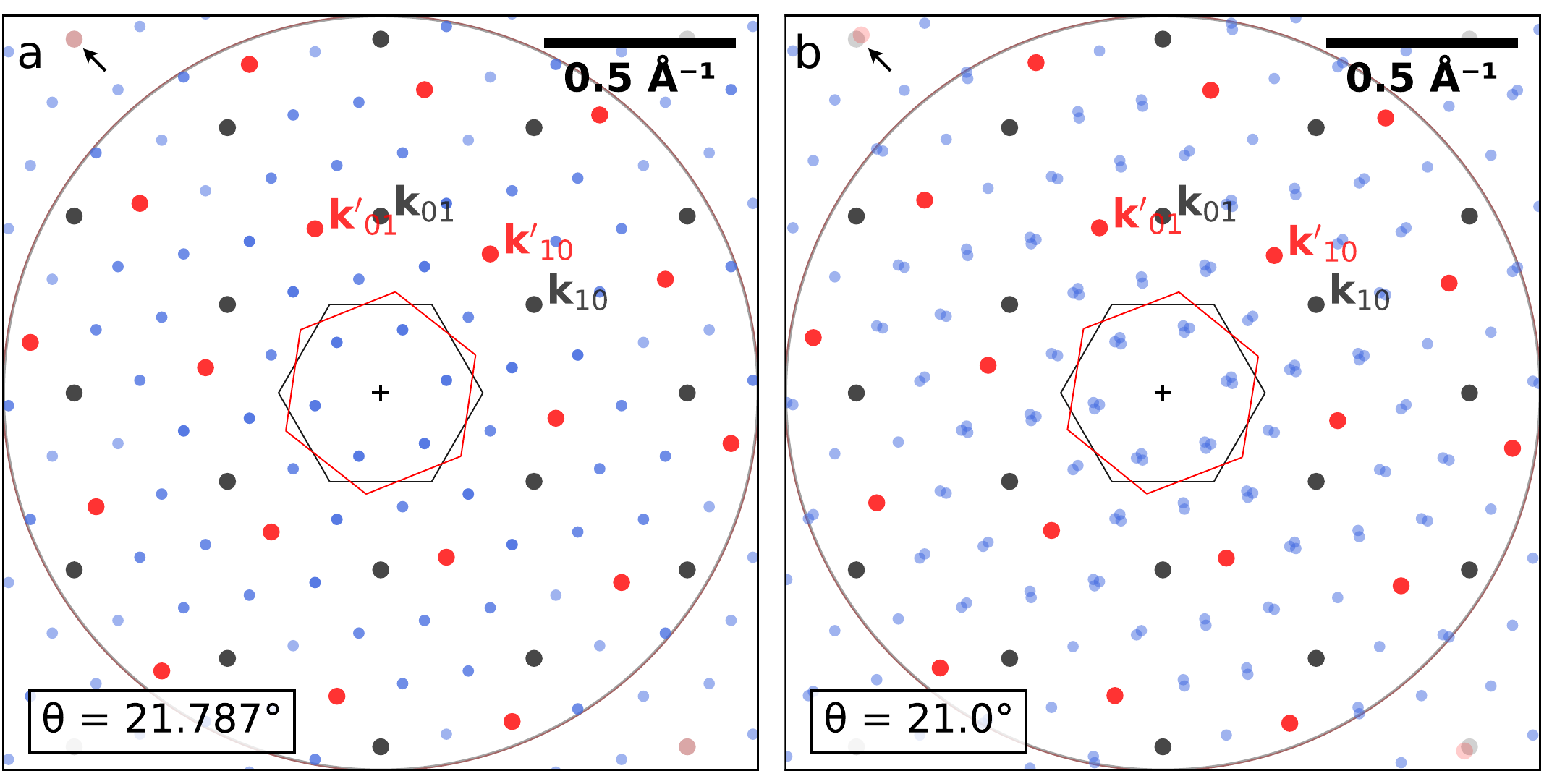}
    \caption{Commensurability and moir\'e wavevectors. (a) Reciprocal lattices of commensurate bilayer graphene twisted by $\theta=21.787^\circ$, (b) incommensurately twisted by $\theta=21.0^\circ$, both for $k_0=k_0'=1.0$~\AA$^{-1}$. The moir\'e wavevectors $\mathbf{M}_{mnpq}$ are in light blue. In the commensurate case, $\mathbf{M}_{mnpq}$ form a regular lattice in reciprocal space, \emph{i.e.}, there are multiple sets of indices $(m,n,p,q)$ for a given moir\'e wavevector $\mathbf{M}$ (the opacity of the light blue dots is increased when multiple $\mathbf{M}$ occupy the same coordinates). In the incommensurate case, the opposite is true; each $\mathbf{M}\in\mathbf{M}_{mnpq}$ has unique indices. In (a) and (b), the black arrows point to $\mathbf{k}_{\bar{2}3}$ and $\mathbf{k}'_{\bar{1}3}$, equal in (a) and dissimilar in (b).}
    \label{fig:s4}
\end{figure*}

Last, we consider an extreme case where the reciprocal cutoff is largely overestimated. Figure~\ref{fig:si_k0}(d) shows the same system with $k_0=k'_0=0.70$~\AA$^{-1}$. The number of predicted $\mathbf{M}_{mnpq}$ is very large (192 wavevectors) due to the large number of reciprocal lattice points included in the calculation. This case is similar to the previous case in Fig.~\ref{fig:si_k0}(c), however the cutoff is so large that the MPWEM predicts $\mathbf{M}_{\bar{2}1\bar{3}0}$ in close proximity to the origin (see arrow in inset). Considering the amplitude assignment scheme which promotes low $|\mathbf{M}|$ moir\'e plane waves ($a_M\propto|\mathbf{M}|^{\eta}$, $\eta<0$), the plane wave associated with $\mathbf{M}_{\bar{2}1\bar{3}0}$ would dwarf other amplitudes and lead to a very poor agreement with the data.

In general, there is no comprehensive method for the \emph{a priori} determination of the reciprocal cutoff. However, the systems successfully modelled using the MPWEM in the main text have comparable cutoff parameters ($k_0\sim 0.3 - 0.5$). More generally, the cutoff parameters are set to include the features present in the FFT, and not beyond. For instance, in the $\alpha$-Bi/MoS$_2$ case, $\mathbf{k'}_{20}$ is visible in the data, but $\mathbf{k'}_{21}$ is not; therefore $\mathbf{k'}_{20}<k'_0<\mathbf{k'}_{21}$. The same approach was used in graphene/WTe$_2$.

To conclude this section, underestimating the cutoff may lead to underestimation of the number of relevant $\mathbf{M}_{mnpq}$ that will model actual moir\'e modulations; on the other hand, overestimating the cutoff can lead to generating additional low-frequency artificial moir\'e modulations which do not exist in the real space data and have no physical origin. If the STM data is available, the optimum cutoff is found by trial-and-error; otherwise the optimal window seems to correspond to $k_0$ slightly larger than $\max{\{|\mathbf{k}_{10}|, |\mathbf{k}_{01}|, |\mathbf{k'}_{10}|, |\mathbf{k'}_{01}| \}}$.


\section{Commensurability}
\label{si:commensurability}

We now evaluate the consequence of commensurability on the structure of the $\mathbf{M}_{mnpq}$ set. We use the simple case of bilayer graphene (each layer has identical symmetry). Figure \ref{fig:s4}(a) shows the commensurate case where $\theta=21.787^\circ$. The commensurability is confirmed by the exactly overlapping reciprocal lattice points: $\mathbf{k}_{\bar{2}3}=\mathbf{k}'_{\bar{1}3}$ (black arrow in the top left, beyond $k_0=k_0'=1.0$~\AA$^{-1}$). The moir\'e wavevectors $\mathbf{M}_{mnpq}$ resulting from this superposition (in blue) form a triangular periodic lattice (like the reciprocal lattices of both graphene layers). Increasing $k_0$ and $k'_0$ leads to more moir\'e wavevectors, but no new coordinates within the cutoff radii are obtained. We can show this rigorously as follows. The commensurate condition is that there exists non trivial $M, N, P, Q\in\mathbb{Z}$ such that:
\begin{equation}
\label{eq:commensuratedef}
\mathbf{k}_{MN}=\mathbf{k}'_{PQ}
\end{equation}
which implies that:
\begin{equation}
\label{eq:commensurate}
\begin{split}
\mathbf{M}_{m+M, n+N, p+P, q+Q} =~&\mathbf{k}'_{p+P, q+Q} - \mathbf{k}_{m+M, n+N}\\
  =~&\mathbf{k}'_{pq}-\mathbf{k}_{mn} + (\mathbf{k}'_{PQ} - \mathbf{k}_{MN})\\
  =~&\mathbf{k}'_{pq}-\mathbf{k}_{mn} = \mathbf{M}_{mnpq},
\end{split}
\end{equation}
which means that a given $\mathbf{M}$ in the set of moir\'e wavevectors has multiple sets of indices $(m,n,p,q)$ which lead to $\mathbf{M}=\mathbf{M}_{mnpq}$.

Figure \ref{fig:s4}(b) shows a very similar bilayer graphene system, with a slight modification of the twist angle ($\theta=21.0^\circ$) such that the superposition is now incommensurate. The location of the $\mathbf{M}_{mnpq}$ is roughly the same, but the wavevectors that shared the same coordinates in the commensurate case in Fig.~\ref{fig:s4}(a) are now separated. Rigorously, the incommensurate case mean that $\forall~(n,m,p,q)\neq (N,M,P,Q): \mathbf{M}_{mnpq}\neq\mathbf{M}_{MNPQ}$ (as always indices $\in\mathbb{Z}$). In the following, we prove by contradiction that it is indeed the case. Suppose there exists $(n, m, p, q) \neq (N, M, P, Q)$ in an incommensurate system such that:
\begin{equation}
\begin{split}
&\mathbf{M}_{mnpq} =\mathbf{M}_{MNPQ}\\
&\Rightarrow \mathbf{k}'_{pq}-\mathbf{k}_{mn} =\mathbf{k}'_{PQ}-\mathbf{k}_{MN}\\
&\Rightarrow \mathbf{k}'_{p-P,q-Q} - \mathbf{k}_{m-M,n-N} = 0
\end{split}
\end{equation}
which is equivalent to the condition of commensurability in eq. (\ref{eq:commensuratedef}), leading to a contradiction.

\section{Computational methods}
\label{si:computationalmethods}

We briefly show the python structure and concepts required to generate a simulated STM image using the MPWEM. Example files implementing the method can be found at {\url{https://github.com/maximelester/mpwem/}}.

\paragraph{Structure} First, it is convenient to introduce the general data structure describing the individual layers $z$, $z'$ and the interaction term $z_M$. Because they are based on a plane wave description, we use a 2D {\fontfamily{lmtt}\selectfont numpy} array containing the two ($n,m$ or $p,q$) indices for either of the non-interacting layers (or four ($m,n,p,q$) indices for the interaction term) as well as the four parameters that describe the 2D plane wave: $k_x, k_y, \varphi$ and $a$ ($k_x, k_y$ are in \AA$^{-1}$ units,  $a$ in \AA~and $\varphi$ in radians). A typical plane wave parameters list is conveniently shown as follows (graphene plane wave parameters as an example):

\begin{verbatim}

       m	 n	|  kx	    ky	    phi	  a	  
      -----------------------------------
       0	-1	| -0.368 -0.287	-0.541	0.196 
       1	-1	|  0.065 -0.462	-0.871	0.196  
       1	 0	|  0.433 -0.175	-0.330	0.196  
       0	 0	| -0.000  0.000  0.000	0.510 
\end{verbatim}
where the $(m,n)=(0, 0)$ term (\emph{i.e.}, the average value) is kept, although will not be included in the calculation of the interaction. The objective is to obtain a plane wave parameters list as above for both substrate (describing $z$) and adsorbate layers (describing $z'$). To do so, there are two separate avenues: (1) fitting existing atomically-resolved STM data, or if unavailable or in unsufficient quality, (2) generation from lattice parameters. Both methods rely on successfully extracting the plane wave parameters via fitting using a standard least-square algorithm, {\fontfamily{lmtt}\selectfont scipy.optimize.curve\_fit}, to fit the image obtained in (1) or (2).

\paragraph{Fitting STM data.} To fit atomically-resolved STM data, it is crucial that the reciprocal lattice vectors $\mathbf{k}_{10}$ and $\mathbf{k}_{01}$ are known, typically this is achieved via FFT analysis (or can be simply derived from real space lattice parameters). A list of plane wave parameters akin the one shown above as an example is generated such that all $\mathbf{k}_{mn}$ are included up to the cut off $k_0$, \emph{i.e.}, $|\mathbf{k}_{mn}|<k_0$. The initial `guess' amplitudes are set to $a_{mn}=1/|\mathbf{k}_{mn}|^2$ and $a_{00} = \langle z \rangle$ (average value). Importantly, the plane wave list excludes \emph{twin} plane wave parameters ($\mathbf{k}_{\bar{m}\bar{n}} = -\mathbf{k}_{mn}$) as they would otherwise lead to a high correlation, impacting the quality and efficiency of the fitting. If the reciprocal lattice base vectors are correctly determined from FFT analysis, the fitting times are of the order of a few seconds.

\paragraph{Generating STM data.} If atomic resolution STM data is unavailable (or of poor quality, \emph{e.g.} resulting from a blunt tip), the STM data can be generated from basic surface crystallography parameters such as $\mathbf{R}_1$, $\mathbf{R}_2$ (which define the lattice unit cell) and additional terms describing the atomic population within the unit cell: {\fontfamily{lmtt}\selectfont atoms} (a list of $(x,y)$ positions in fractional coordinates), {\fontfamily{lmtt}\selectfont radii} (a list of radii for each element of {\fontfamily{lmtt}\selectfont atoms}, {\fontfamily{lmtt}\selectfont weights} (a list of relative weights (or heights) for each element of {\fontfamily{lmtt}\selectfont atoms}), and lastly a global translation term {\fontfamily{lmtt}\selectfont offset} (a 2D vector describing the translation from the real space origin). For each atom of the unit cell, we sum cosine terms (laterally offset based on the coordinates contained in {\fontfamily{lmtt}\selectfont atom} and global translation in {\fontfamily{lmtt}\selectfont offset}) and raise them individually to a power $\zeta$ in order to \emph{shape} the cosine terms as desired by the radius contained in the {\fontfamily{lmtt}\selectfont radii} list. 

\paragraph{Optimum parameters from STM image.} In order to find the optimum parameters $\mu, \tau, \eta$ and $a_0$, we proceed to a series of fitting (still using an algorithm based on {\fontfamily{lmtt}\selectfont scipy.optimize.curve\_fit}. The image is separated into the substrate, adsorbate and moir\'e terms, for which we have separate fitting. The amplitudes of fitting (not normalized) are $\tilde{a}_i$, $\tilde{a}'_i$ and $\tilde{a}_{Mi}$ for the substrate, adsorbate and moir\'e terms respectively. It follows that:

\begin{equation}
\tau = \frac{1}{1+\frac{\sum_i \tilde{a}_i}{\sum_i \tilde{a'}_i}}
\end{equation}

\begin{equation}
\mu = \frac{1}{1+\frac{1}{\tau}\frac{\sum_i \tilde{a}_i +\sum_j \tilde{a'}_j}{\sum_i \tilde{a}_{Mi}}}
\end{equation}

\begin{equation}
a_0 = \frac{1}{\mu}\sum_i \tilde{a}_{Mi}
\end{equation}

To determine $\eta$, the experimental amplitudes $a_M$ are plot as a function of $|\mathbf{M}_{mnpq}|$ and fit using {\fontfamily{lmtt}\selectfont scipy.optimize.curve\_fit}, this time using a power law function of the form $a = c|\mathbf{M}|^\eta$ ($c\in\mathbb{R}$) as performed in section SI \ref{si:amplitudes}.

\end{document}